\documentclass[preprint, authoryear, 3p]{elsarticle}

\usepackage{hyperref}
\pdfoutput=1
\usepackage{hyperref}
\usepackage{array} 
\usepackage{algorithm}
\usepackage{algcompatible} 
\usepackage{multirow}
\usepackage{algpseudocode}
\usepackage{xcolor}
\usepackage{booktabs}
\usepackage{stfloats} 
\usepackage{pgfplots, pgfplotstable}

\usepgfplotslibrary{groupplots}
\pgfplotsset{compat=newest}
\pgfplotsset{plot coordinates/math parser=false}
\usepackage{pdflscape}
\usepackage{adjustbox}

\usepackage{verbatim}
\usepackage{graphicx}
\usepackage{amsmath}
\usepackage{setspace}
\usepackage{rotating}
\usepackage{placeins}
\usepackage{afterpage}
\usepackage{appendix}

\usepackage{amsfonts}
\usepackage{mathtools} 

\usepackage{amsthm}

\usepackage{tikz}
\usetikzlibrary{matrix}
\usepackage{siunitx}
\usetikzlibrary{arrows,calc}
\usepackage{relsize}

\usepackage{threeparttable}
\usepackage{longtable}
\usepackage{caption}
\usepackage{subcaption}

\makeatletter
\def\els@aparagraph[#1]#2{\elsparagraph[#1]{#2 }}
\def\els@bparagraph#1{\elsparagraph*{#1\@addpunct{.}}}
\def\ps@pprintTitle{%
\let\@oddhead\@empty
\let\@evenhead\@empty
\def\@oddfoot{}%
\let\@evenfoot\@oddfoot}
\makeatother

\begin{document}

\begin{frontmatter}

\title{Adaptive Large Neighborhood Search for Vehicle Routing Problems with Transshipment Facilities Arising in City Logistics}

\author{Christian Friedrich*}\cortext[cor1]{Corresponding author: chris.friedrich@gmx.net}
\author{Ralf Elbert}
\address{Chair of Management and Logistics, Department of Law and Economics, Technical University of Darmstadt, Hochschulstr.~1, 64289 Darmstadt, Germany}

\begin{abstract}
In this paper, we investigate vehicle routing problems with third-party transshipment facilities that arise in the context of city logistics. Contrary to classical vehicle routing problems, where each customer request is delivered directly to its destination, the problems considered in this paper feature the alternative possibility of delivering customer requests to third-party transshipment facilities, such as urban consolidation centers, for a fee. We present an adaptive large neighborhood search with an embedded random variable neighborhood descent as a local search component and a set-partitioning problem for the recombination of routes to solve various versions of the problem. Thereby, we consider location-dependent time windows as well as heterogeneous fleets and propose several new procedures that consider transshipment facilities within the components of our adaptive large neighborhood search. The proposed method is tested on benchmark instances from the literature as well as newly created benchmark instances. It shows promising results, leading to multiple improvements over existing algorithms from the literature. Moreover, a real-world study is presented to gain managerial insights on the impact of transshipment fees, order size, and heterogeneous fleets on the transshipment decisions.
\end{abstract}


\end{frontmatter}


\section{Introduction}\label{sec:introduction}

City logistics and last-mile deliveries give rise to numerous challenges for logistics service providers (LSPs), such as increased public attention to sustainability \citep{Savelsbergh.2016}, urban access restrictions (\cite{Ville.2013}, \cite{Elbert.2018}), and growing delivery volumes \citep{EU.2019}. A common approach to address these challenges is the consolidation of shipments at transshipment facilities to increase the efficiency of urban freight transport. In this context, Urban Consolidation Centers (UCCs) are a widely studied concept in city logistics (\cite{Bjorklund.2018}, \cite{Lagorio.2016}). UCCs can be defined as logistics transshipment facilities in the proximity of an urban area that consolidate urban freight transport across companies \citep{Allen.2012}. Although numerous studies have been published on UCCs and several UCCs have been implemented in cities across the world, except for a few cases (e.g. \cite{Baldacci.2017}), the decision on whether to outsource individual shipments to third-party transshipment facilities, such as UCCs, has rarely been included in vehicle routing problems.

Motivated by this lack of research, and to bridge the gap towards this real-world problem in city logistics, in this paper we expand the research on the \textit{vehicle routing problem with transshipment facilities} (VRPTF)\citep{Baldacci.2017} by including location-dependent time windows and heterogeneous fleets. As a generalization, this problem can be described as a vehicle routing problem that includes selecting delivery locations for customer requests from among multiple possible delivery locations, each with the possibility of different transshipment costs and time windows. These transshipment facilities could be, for example, UCCs serving an urban area, or regular third-party logistics service provider facilities.Aligning with problem notations from the literature, we refer to our new problem variants of the VRPTF as the \textit{vehicle routing problem with time windows and transshipment facilities} (VRPTWTF) and the \textit{fleet size and mix vehicle routing problem with time windows and transshipment facilities} (FSMTWTF).

The purpose of this paper is twofold. First, it contributes to filling the literature gap in that it extends the body of literature on vehicle routing problems with transshipment facilities by adding time windows and heterogeneous fleets as additional attributes. Second, it presents an efficient meta-heuristic to solve these problems, which is also tested on related problems. Additionally, our computational results show that the solutions of existing VRPTF instances from literature can be improved by 0.68~\% on average.

The outline of this paper is as follows: Section~\ref{sec:literature} gives an overview of the related literature. In Section \ref{sec:problem} the problem is described. Section \ref{sec:alns} presents the details of our adaptive large neighborhood search (ALNS). Section \ref{sec:results} describes how we tuned our algorithm and reports computational results on different sets of problems. Finally, Section \ref{sec:conclusion} provides concluding remarks and pointers for further research.

\section{Related Literature}\label{sec:literature}
Considering the selection of delivery locations such as transshipment facilities in vehicle routing is a generalization of the classical capacitated vehicle routing problem. To the best of our knowledge, only a few papers consider the selection of delivery locations or third-party transshipment facilities in vehicle routing. Most notably, \cite{Baldacci.2017} formulate the \textit{vehicle routing problem with transshipment facilities} (VRPTF) where customers can either be served directly from a central depot or by using a transshipment facility for a fee. In their paper, they describe various valid inequalities for the VRPTF and provide a mathematical formulation of the problem. Furthermore, they derive lower bounds from a set-partitioning (SP) based formulation of the VRPTF.

\cite{Alcaraz.2019} also address the topic of using transshipment facilities for last-mile deliveries. Similar to the problem studied by \cite{Baldacci.2017}, they assume that some customer requests can be delivered directly or to a transshipment facility, from where a third-party subcontractor will perform the final last-mile deliveries for a fee. They study a rich vehicle routing problem that focuses on long-haul transport between regions and features several additional characteristics, such as driving hour regulations, incompatibility among goods, heterogeneous vehicles, and time windows. They present a construction heuristic method that incorporates these characteristics and adapt various classical improvement heuristics from literature to solve their problem.

Similarly, \cite{Sitek.2019} study a so-called \textit{capacitated vehicle routing problem with pick-up and alternative delivery} (CVRPPAD). They assume that customer requests can have alternative delivery locations such as a parcel locker or postal outlets and consider time windows and heterogeneous fleets. To solve their problem they propose an exact solution based on mathematical programming and a heuristics solution for larger problem instances. In their heuristics solution method, customer requests are assigned to routes based on a set of rules, and subsequently, a traveling salesman problem is solved for each of these routes.

\cite{Dumez.2021_trb} also focus on different delivery locations for customers and recently introduced the \textit{vehicle routing problem with delivery options} (VRPDO). In this problem, each customer can specify a number of delivery options along with a preference value. Furthermore, each delivery option can have a time window and capacity. Contrary to the VRPTF, no additional costs are associated with the shared delivery options and the focus is placed on satisfying the customer preference levels and shared  location capacity constraints. In order to solve the problem, the authors present a large neighborhood search (LNS). Extending their work on the VRPDO, \cite{Dumez.2021} introduce an adapted Balas-Simonetti neighborhood to their LNS to further improve the solution quality.  Moreover, \cite{Tilk.2020}, also address the VRPDO and present a branch-price-and-cut algorithm that solves instances with up to 50 customers and 100 options to optimality.

Focusing on a problem with home deliveries and shared delivery options, \cite{Mancini.2021} propose two matheuristic-based solution methods. Similar to the VRPTF, and in contrast to the VRPDO, their problem variant introduces a fee that has to be paid for each delivery assigned to a shared delivery location.

\cite{Zhou.2018} consider a \textit{two-echelon vehicle routing problem} (2E-VRP), where deliveries are routed through an intermediate capacitated satellite depot. Similar to \cite{Dumez.2021_trb} and \cite{Sitek.2019}, they assume that customers can provide different delivery options, such as receiving their parcels at their home or picking them up at an intermediate pick up facility. To solve this problem, \cite{Zhou.2018} propose a hybrid multi-population genetic algorithm that is tested on real-world and artificial instances. Although the 2E-VRP is generally similar to the VRPTF, \cite{Baldacci.2017} point out two significant differences between the two problems. First of all, the facilities and satellites in the 2E-VRP are assumed to be owned and operated by the same LSPs as the main depot. In the VRPTF, the facilities are operated by third-party providers that charge the LSP using its service a fee for conducting the last-mile deliveries to the customers. As a consequence, the vehicle routing from the transshipment facility to the customers is not part of the VRPTF. Second, in the 2E-VRP each customer request is generally routed through an intermediate facility, while in the VRPTF the customer requests can be either transshipped at a facility or delivered directly.

Two other closely related problems to the VRPTF that also include multiple delivery locations are the \textit{vehicle routing problem with roaming delivery locations} (VRPRDL) and the very similar \textit{vehicle routing problem with home and roaming delivery locations} (VRPHRDL). \cite{Reyes.2017} define the VRPRDL as a routing problem where customer shipments are delivered to the trunk of their cars. Thereby, the cars are parked in different locations at different times, based on a known schedule. However, contrary to the VRPTF, no additional costs are associated with the different delivery locations and customers do not share common locations, such as a transshipment facility. \cite{Reyes.2017} present a construction heuristic and an LNS-based improvement heuristic for solving this problem. The VRPHRDL introduced by \cite{Ozbaygin.2017} extends the VRPRDL to allow shipments to be delivered to customers' homes during the entire planning period or, based on a predefined schedule, to the trunk of their cars. In order to solve both the VRPRDL and VRPHRDL, \cite{Ozbaygin.2017} developed a branch-and-price algorithm. Later, \cite{Dumez.2021_trb}, \cite{Dumez.2021}, as well as \cite{Tilk.2020} also study both problems to validate their algorithms for the VRPDO, leading to new best solutions and large time improvements on the instances of \cite{Ozbaygin.2017} and \cite{Reyes.2017}.

Although transshipment facilities such as UCCs have been a topic in city logistics for a long time, dating back to as early as the 1970s \citep{Allen.2012}, only a few studies focus on whether to transship goods at a third-party transshipment facility. To the best of our knowledge, the studies on UCCs usually assume that only a single transshipment facility (UCC) exists, and that either all or no deliveries for a city are transshipped at the UCC. In reality, however, more than one transshipment facility can exist. Besides, as indicated in a survey among LSPs by \cite{Stathopoulos.2012}, it might be preferable to transship only some customer requests (e.g. requests with narrow time windows), while other customer requests are served directly. Moreover, the transshipment decision is often modeled to be periodical, depending on either the past performance or cost estimates (e.g. \cite{Firdausiyah.2019}, \cite{vanHeeswijk.2019}, \cite{vanDuin.2012}). 

The literature review underlines that the literature on selecting delivery locations in the context of vehicle routing is still scarce. Especially third-party transshipment facilities in connection with heterogeneous vehicles have been little studied so far. Hence we aim to expand the literature in this regard.

\section{Problem Description}\label{sec:problem}

In our paper, we consider different problem variants of the vehicle routing problem with transshipment facilities (VRPTF). Table~\ref{tab:problems} gives an overview of the considered problem variants and their properties. The VRPTWTF extends the VRPTF by including time window constraints, while the FSMTWTF is based on the \textit{fleet size and mix vehicle routing problem with time windows} (FSMTW) \citep{Liu.1999} and extends the VRPTWTF by adding fleet size and mix decisions. For both problem variants, we distinguish between two cost functions consisting of different vehicle cost components. In the following, we describe the variant of the FSMTWTF with a rich cost function, abbreviated FSMTWTF-R, which encompasses the properties of the other problem variants studied.

\begin{table}[htb]
\footnotesize
\centering
\caption{Overview of considered problem variants and their aspects.}\label{tab:problems}
{\begin{tabular}{@{\extracolsep{2pt}}l cl ccc c}
\toprule
\multirow{2}{2.2cm}{Problem variant} & \multirow{2}{1.8cm}{Time windows} &\multicolumn{2}{c}{Fleet} &\multicolumn{3}{c}{Vehicle cost}\\
\cline{3-4}\cline{5-7}\rule{0pt}{2.5ex}
&& homogeneous& heterogeneous & fixed & distance & duration\\
\midrule
VRPTF &  & \checkmark & & &\checkmark\\
VRPTWTF & \checkmark & \checkmark&  & & \checkmark &\\
FSMTWTF & \checkmark & & \checkmark & \checkmark & \checkmark &\\
VRPTWTF-R & \checkmark &\checkmark  & & \checkmark & \checkmark &\checkmark\\
FSMTWTF-R & \checkmark & & \checkmark & \checkmark & \checkmark &\checkmark\\
\bottomrule
\end{tabular}}
\end{table}

Expanding the notation of the VRPTF from \cite{Baldacci.2017}, the FSMTWTF-R considered in this paper can be described as follows. Let $G=(V,A)$ be a complete directed graph, where $V = \{v_0\} \cup V'$ is the set of vertices and $A$ the set of arcs. The set of vertices $V'$ is partitioned into $V'=\{V_C, V_F\}$, where $V_C = \{v_1,\dots,v_{n_C}\}$ is the set of customer locations and $V_F = \{v_{n_C+1},\dots,v_{n_C+n_F}\}$ the set of transshipment facilities. Vertex $v_0$ represents the depot, where goods to be delivered are stored and vehicle routes originate. 

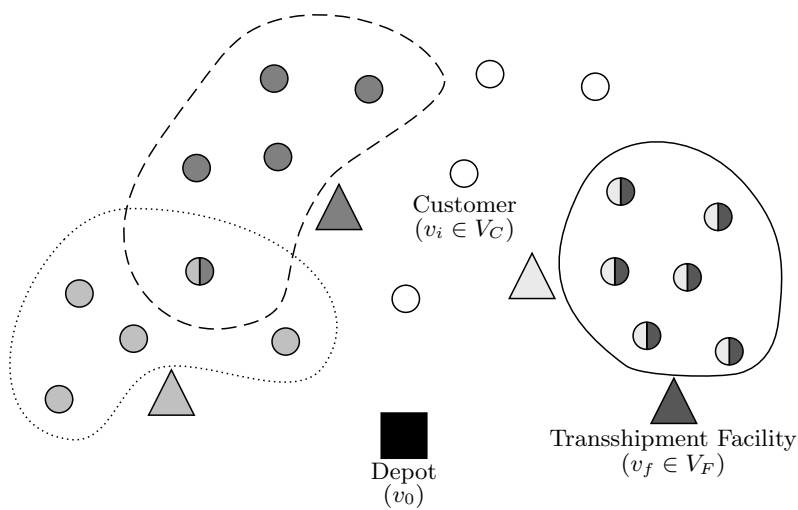
\begin{figure}[htb]
\centering
\begin{tikzpicture}[x=0.60mm, y=0.60mm, inner xsep=0pt, inner ysep=0pt, outer xsep=0pt, outer ysep=0pt]
\path[line width=0mm] (65.25,10.82) rectangle +(175.91,115.33);
\definecolor{L}{rgb}{0,0,0}
\definecolor{F}{rgb}{0,0,0}
\path[line width=0.20mm, draw=L, fill=F] (150.01,24.82) rectangle +(10.00,10.00);
\definecolor{F}{rgb}{0.502,0.502,0.502}
\path[line width=0.20mm, draw=L, fill=F] (126.54,108.72) circle (1.80mm);
\path[line width=0.20mm, draw=L] (230.12,84.96) .. controls (219.24,97.90) and (200.53,98.34) .. (193.63,84.34) .. controls (185.60,68.04) and (187.54,57.14) .. (203.64,45.30) .. controls (207.08,42.77) and (224.54,42.49) .. (228.61,43.64) .. controls (233.66,45.07) and (236.42,49.35) .. (237.08,54.78) .. controls (238.64,67.55) and (238.35,75.16) .. (230.12,84.96) -- cycle;
\path[line width=0.20mm, draw=L, dash pattern=on 0.20mm off 0.50mm] (80.55,73.25) .. controls (65.98,58.03) and (62.81,31.24) .. (81.84,28.96) .. controls (91.32,27.82) and (92.42,42.80) .. (101.63,44.89) .. controls (111.74,47.18) and (131.11,36.42) .. (137.25,45.04) .. controls (148.72,61.14) and (126.16,80.03) .. (95.47,80.00) .. controls (95.35,80.00) and (95.24,80.00) .. (95.12,80.00) .. controls (89.55,80.01) and (84.41,77.28) .. (80.55,73.25) -- cycle;
\path[line width=0.20mm, draw=L, dash pattern=on 2.00mm off 1.00mm] (163.64,109.43) .. controls (157.26,122.66) and (130.95,127.42) .. (118.96,117.57) .. controls (115.12,114.42) and (112.34,110.24) .. (109.40,106.25) .. controls (100.96,94.79) and (90.83,82.76) .. (94.26,69.06) .. controls (96.83,58.80) and (106.57,51.93) .. (116.90,53.63) .. controls (136.76,56.89) and (123.59,66.56) .. (138.43,86.41) .. controls (142.68,92.08) and (166.72,103.03) .. (163.64,109.43) -- cycle;
\draw(155.02,20.27) node[anchor=base]{\fontsize{9}{10.24}\selectfont Depot};
\definecolor{F}{rgb}{0.753,0.753,0.753}
\path[line width=0.20mm, draw=L, fill=F] (99.00,34.35) -- (104.00,44.35) -- (109.00,34.35) -- cycle;
\definecolor{F}{rgb}{0.502,0.502,0.502}
\path[line width=0.20mm, draw=L, fill=F] (135.74,75.36) -- (140.74,85.36) -- (145.74,75.36) -- cycle;
\definecolor{F}{rgb}{0.918,0.918,0.918}
\path[line width=0.20mm, draw=L, fill=F] (178.09,60.12) -- (183.09,70.12) -- (188.09,60.12) -- cycle;
\definecolor{F}{rgb}{0.341,0.341,0.341}
\path[line width=0.20mm, draw=L, fill=F] (209.19,32.52) -- (214.19,42.52) -- (219.19,32.52) -- cycle;
\definecolor{F}{rgb}{0.502,0.502,0.502}
\path[line width=0.20mm, draw=L, fill=F] (109.52,88.95) circle (1.80mm);
\path[line width=0.20mm, draw=L, fill=F] (147.32,106.36) circle (1.80mm);
\definecolor{F}{rgb}{0.753,0.753,0.753}
\path[line width=0.20mm, draw=L, fill=F] (83.82,61.23) circle (1.80mm);
\path[line width=0.20mm, draw=L, fill=F] (79.36,37.87) circle (1.80mm);
\path[line width=0.20mm, draw=L, fill=F] (129.06,50.61) circle (1.80mm);
\path[line width=0.20mm, draw=L, fill=F] (95.72,51.25) circle (1.80mm);
\definecolor{F}{rgb}{0.502,0.502,0.502}
\path[line width=0.20mm, draw=L, fill=F] (127.36,91.39) circle (1.80mm);
\draw(214.06,27.50) node[anchor=base]{\fontsize{9}{10.24}\selectfont Transshipment Facility};
\draw(168.32,79.23) node[anchor=base]{\fontsize{9}{10.24}\selectfont Customer};
\definecolor{F}{rgb}{1,1,1}
\path[line width=0.20mm, draw=L, fill=F] (168.18,87.73) circle (1.80mm);
\path[line width=0.20mm, draw=L, fill=F] (155.40,60.06) circle (1.80mm);
\path[line width=0.20mm, draw=L, fill=F] (196.96,106.93) circle (1.80mm);
\path[line width=0.20mm, draw=L, fill=F] (173.87,109.71) circle (1.80mm);
\draw(214.06,21.99) node[anchor=base]{\fontsize{9}{10.24}\selectfont $(v_f \in V_F)$};
\draw(168.32,74.05) node[anchor=base]{\fontsize{9}{10.24}\selectfont $(v_i \in V_C)$};
\draw(155.02,15.23) node[anchor=base]{\fontsize{9}{10.24}\selectfont $(v_0)$};
\definecolor{F}{rgb}{0.341,0.341,0.341}
\path[line width=0.20mm, draw=L, fill=F] (217.18,61.81) arc (-90:90:1.80mm) -- cycle;
\definecolor{F}{rgb}{0.918,0.918,0.918}
\path[line width=0.20mm, draw=L, fill=F] (217.13,67.81) arc (90:270:1.80mm) -- cycle;
\definecolor{F}{rgb}{0.341,0.341,0.341}
\path[line width=0.20mm, draw=L, fill=F] (202.55,80.75) arc (-90:90:1.80mm) -- cycle;
\definecolor{F}{rgb}{0.918,0.918,0.918}
\path[line width=0.20mm, draw=L, fill=F] (202.50,86.75) arc (90:270:1.80mm) -- cycle;
\definecolor{F}{rgb}{0.341,0.341,0.341}
\path[line width=0.20mm, draw=L, fill=F] (201.26,63.10) arc (-90:90:1.80mm) -- cycle;
\definecolor{F}{rgb}{0.918,0.918,0.918}
\path[line width=0.20mm, draw=L, fill=F] (201.21,69.10) arc (90:270:1.80mm) -- cycle;
\definecolor{F}{rgb}{0.341,0.341,0.341}
\path[line width=0.20mm, draw=L, fill=F] (223.92,75.15) arc (-90:90:1.80mm) -- cycle;
\definecolor{F}{rgb}{0.918,0.918,0.918}
\path[line width=0.20mm, draw=L, fill=F] (223.88,81.15) arc (90:270:1.80mm) -- cycle;
\definecolor{F}{rgb}{0.341,0.341,0.341}
\path[line width=0.20mm, draw=L, fill=F] (226.36,45.45) arc (-90:90:1.80mm) -- cycle;
\definecolor{F}{rgb}{0.918,0.918,0.918}
\path[line width=0.20mm, draw=L, fill=F] (226.32,51.45) arc (90:270:1.80mm) -- cycle;
\definecolor{F}{rgb}{0.341,0.341,0.341}
\path[line width=0.20mm, draw=L, fill=F] (208.29,48.89) arc (-90:90:1.80mm) -- cycle;
\definecolor{F}{rgb}{0.918,0.918,0.918}
\path[line width=0.20mm, draw=L, fill=F] (208.24,54.89) arc (90:270:1.80mm) -- cycle;
\definecolor{F}{rgb}{0.502,0.502,0.502}
\path[line width=0.20mm, draw=L, fill=F] (110.23,63.13) arc (-90:90:1.80mm) -- cycle;
\definecolor{F}{rgb}{0.753,0.753,0.753}
\path[line width=0.20mm, draw=L, fill=F] (110.18,69.13) arc (90:270:1.80mm) -- cycle;
\end{tikzpicture}

\caption{Example of a VRPTF instance with four transshipment facilities. Each transshipment facility (triangles) has a number of customer locations that it can serve, indicated by the grey scale of each locations' circle and the surrounding polygon. Some customer requests, denoted as white circles, can only be served directly and may not be transshipped.}\label{fig:problem example}
\end{figure}

Each customer location $v_i \in V_C$ is associated with a single customer request $cr_i$. These requests are characterized by a non-negative demand of $q_i$ units to be delivered from the depot $v_0$ and a non-negative service time $t^s_i$. Each customer request can be delivered directly to the associated customer $v_i$ or, if applicable, to a transshipment facility selected from its set $F_i \subseteq V_F$ of possible transshipment facilities. Figure~\ref{fig:problem example} gives an example of a vehicle routing problem with four transshipment facilities, where customer requests can be assigned to a single facility, multiple facilities, or no transshipment facility at all.

The selected delivery location of a customer request $cr_i$ is denoted as $cr_i^l \in F_i \cup \{v_i\}$. Each location from $V'$ can have a delivery time window specified by the earliest and latest time $tw_i^S$ and $tw_{i}^E$ ($0\leq tw_i^S < tw_i^E <\infty$) to be visited. In order to account for possible time savings when customer requests are consolidated and transshipped at a transshipment facility, a preparation time $t^p_i$ is assigned to each location similar to the approach of \cite{Dumez.2021_trb}. This preparation time represents, for example, the time needed for parking and administrative tasks (e.g. handling delivery documents) at the locations that occur regardless of the quantity being unloaded. In contrast to this, the service time $t^s_i$ is location independent and represents the time needed for unloading the goods of a customer request $cr_i$. The arc set $A$ represents the links between the vertices, where $d_{ij}$ is the distance between location $v_i$ and $v_j$ and $t_{ij}$ the transport time. For distinct locations the transport time $t_{ij}$ includes the location-dependent preparation $t^p_i$. 

The fleet originating from the depot is composed of $n_{Vehicles}$ different vehicle types, with $M =\{1,\dots,\allowbreak n_{Vehicles}\}$. For each type $k \in M$, there are $m_k$ available vehicles, each with a capacity $Q_k$. Additionally, fixed costs $c_k^f$ as well as distance-dependent $c_k^d$ and time-dependent costs $c_k^t$ can be assigned to each vehicle type. 

We define a route $r^k$ as a vehicle of type $k$ starting from and ending at the depot $v_0$ and visiting some locations to fulfill customer requests. Each solution $s$ to our problem consists of a set of routes $R$. A route is feasible if the total load of the route, computed as the sum of demand of all customer requests assigned to it, does not exceed the vehicle capacity of the vehicle. Furthermore, for each visited location, and the customer requests assigned to it, the time window constraints must be fulfilled.

The costs $c_r$ of a route $r^k$ consist of four vehicle-dependent cost components. First, the distance-based costs equal the total distance traveled multiplied with the distance-based cost factor $c_k^d$ of the vehicle type. Second, the duration-based costs equal the total operation time to perform the route, including times for traveling, unloading goods at the locations, and possible waiting times due to time windows, multiplied with the duration-dependent cost factor $c_k^t$ of the vehicle type. To minimize the waiting times, the departure time of each vehicle from the depot can be scheduled any time during the work day. Third, the fixed cost $c_k^f$ of the selected vehicle type is incurred when the vehicle is assigned to a tour. Fourth, the sum of transshipment costs for transshipping customer requests at the third-party transshipment facilities. Different transshipment cost functions can be selected for this and included in the objective function. For example, following \cite{Baldacci.2017}, the cost of transshipping customer requests can be based on the distance between the location of the customer and the selected transshipment facility. Alternatively, the cost for assigning a transshipment facility, such as a UCC, can also be based on the quantity $q_i$ of the customer request $cr_i$. This is in line with several real-world UCCs, where the prices usually solely depend on the number of pallets and parcels to be transshipped \citep{Janjevic.2017}. In practice, however, in order to be cost-efficient UCCs usually have a predefined service area, such as a city center, in which they operate (similar to the example in Figure~\ref{fig:problem example}).

In summary, the objective of the vehicle routing problem with transshipment facilities is to determine the assignment of customer requests to transshipment facilities and vehicles as well as a set of routes that minimizes the sum of route costs and does not violate the constraints of the vehicles, customer requests, and locations.

\section{Adaptive Large Neighborhood Search for Vehicle Routing Problems with Transshipment Facilities}\label{sec:alns}
This section describes our adaptive large neighborhood search (ALNS) with an embedded local search for solving the VRPTF and its extensions. ALNS is a common meta-heuristics for solving vehicle routing problems and has been introduced by \cite{Ropke.2006}, who initially applied it to the pickup and delivery problem with time windows. In contrast to classical neighborhoods for VRP, it is characterized by large moves that are performed by removal and insertion procedures. This underlying principle of removing and inserting customer requests, also called ruin and create, has been present in previous research articles. \cite{Shaw.1998} presented the large neighborhood search (LNS), where routes are ruined and subsequently repaired by inserting the unassigned customer requests. Later, \cite{Schrimpf.2000} used the term \textit{ruin and create} for a similar concept to solve several routing problems.

\subsection{Algorithm Outline}

The overall framework of our algorithm to solve the different variants of the VRPTF is reported in Algorithm~\ref{alg:algorithm_outline}. First, we initialize the parameters for the ALNS, such as the stop-criterion, initial weights of the removal and insertion operators, and the conditions for executing the embedded local search procedure and set-partitioning problem (SPP). Second, we create an initial solution by using a regret-2 insertion procedure (see \ref{sec:insertion}). After creating an initial solution, in each iteration of our ALNS, a copy $s'$ of the incumbent solution $s_t$ is destroyed using a removal procedure and subsequently repaired using an insertion procedure (lines 10 and 12). Both removal and insertion procedures are selected randomly based on a roulette wheel selection. Following these two steps, a local search procedure that includes new neighborhoods to modify the locations of customer requests can be invoked with probability $p_{LS}$ to improve the current solution $s'$ (lines 13-15). After the solution improvement phase, we check whether to store the feasible routes of the newly generated solution into the set of routes $R^{pool}$ (line 16). If we find a route that serves the same customers as an existing route in $R^{pool}$ using the same vehicle type with lower costs, we replace the corresponding route. 

Following this, the new solution $s'$ is accepted as the current solution depending on an acceptance criterion (lines 17-18). In case a new best solution has been found, the best solution $s_{best}$ is updated as well (lines 19-20). Every $\eta^{SP}$ iteration of the algorithm the routes of the pool $R^{pool}$ are used to solve a set-partitioning problem and the incumbent solution $s_t$ is updated. After each call to the SPP, $R^{pool}$ is emptied to avoid it becoming too large (lines 24-25). Furthermore, for every $\eta^{update}$ iteration we update the weights (selection probabilities) of the removal and insertion procedures (lines 27-28). Following \cite{Ropke.2006b}, we use an adaptive weight adjustment for the removal and insertion procedures that updates the weight of each procedure after $\eta^{update}$ consecutive iterations (segment). The weights of each procedure $w_{ij}$ are calculated by using the weights of the previous segment and scores obtained during the previous segment as shown in Equation~\ref{eq:weight_update_eq}. The weight $w_{i,j+1}$ of an procedure $i$ in segment $j+1$ is calculated by 

\begin{equation} \label{eq:weight_update_eq}
w_{i,j+1} = w_{ij}(1-\lambda)+\lambda \frac{\pi_{i}}{\theta_{i}}
\end{equation}

where $\pi_{i}$ is the current score of procedure $i$ and $\theta_{i}$ the number of times that the procedure $i$ has been invoked during the last segment. The factor $\lambda$ controls how strongly the procedure's historical performance from previous segments is considered during the calculation. To update the score $\pi_{i}$ each time an insertion or removal procedure has been applied, the reward parameters $\nu_1, \nu_2$, and $\nu_3$ are used to update the procedure's score, as shown in Equation \ref{eq:weight_update_cases}.

\begin{equation}\label{eq:weight_update_cases}
\pi_{i}^{k+1} = \pi_{i}^{k} +
 \begin{cases}
	\nu_1 & \quad\text{if a new best solution has been found} \\
	\nu_2 & \quad\text{if the found solution is better than the incumbent solution}\\
	\nu_3 & \quad\text{if the found solution is accepted but worse than the incumbent solution}
\end{cases}
\end{equation}

By using an adaptive weight adjustment, the selection probabilities of the removal and insertion procedures that previously led to promising solutions are increased, while the selection probabilities of unpromising procedures are reduced.

Lastly, our algorithm stops when either a maximum number of overall iterations $\eta^{max}$ or a maximum number of iterations without improvement $\eta^{max}_{noi}$ has been reached.

\begin{algorithm}[htb]
\begin{footnotesize}
	\caption{Overview of the ALNS algorithm.}\label{alg:algorithm_outline}
	\begin{algorithmic}[1]
\Function{ALNS}{}
	\State $initializeParameters()$
	\State $s_{t}\gets generateInitialSolution()$
	\State $s_{best}\gets s_{t}$ 
	
	\State $k\gets 0$	
	\While{$k<\eta^{max}$ \textbf{and} $k-k_{imp}<\eta^{max}_{noi}$} 
		\State $s' \gets s_t$
		\State Randomly select a removal operator. Draw $\delta$ customers to remove from $s'$.
		\State $s' \gets $ applyRemoval($s',\delta$)
		\State Randomly select an insertion operator to reinsert customers.
		\State $s' \gets $ applyInsertion($s'$)

		\If{random() $< p_{LS}$}
			\State $s'\gets$ LS($s'$) 
		\EndIf	
		\State store the routes of $s'$ in $R^{pool}$
		\If{acceptanceCriteria($s'$,$s_t,s_{best}$)}
			\State $s_{t}\gets s'$ 
			\If{$f(s')<f(s_{best})$}
				\State $s_{best}\gets s'$
			\EndIf
		\EndIf
		\If{\textbf{modulo}($k,\eta^{SP}$)$=0$}
			\State $s_t \gets $ callSetPartitioningProblem($R^{pool}$)
			\State $R^{pool} \gets \emptyset$ 
		\EndIf	
		\If{\textbf{modulo}($k,\eta^{update}$)$=0$}
			\State setNewSelectionScores()
		\EndIf	
		\State $k\gets k+1$	
	\EndWhile
	\EndFunction
	\end{algorithmic}
	\end{footnotesize}
\end{algorithm}

\subsection{Search Space and Objective Function}\label{sec:search_space_and_obj}

In order to allow for infeasible solutions regarding time window constraints during the search, both the removal and insertion procedures, as well as the embedded local search procedure, use the \textit{return in time} relaxation scheme from \cite{Nagata.2010}. In this relaxation scheme, each time a vehicle would arrive late at a location (after the end of the location's time window) it is assumed that the vehicle can travel back in time to arrive at the end of the time window. The amount of time traveled back in time is thereby denoted as \textit{time warp} \citep{Vidal.2015} and used to penalize the objective function. Hence, similar to the approaches of \cite{Olivera.2007} and \cite{Francois.2016}, we use an augmented objective function for evaluating a solution $s'$:

\begin{equation}\label{eq:objective}
f^{mod}(s') = f(s') + \alpha^p  \cdot TW(s')
\end{equation}

where $f(s')$ is the total operating cost of the solution including fixed and variable costs, $TW(s')$ denotes the sum of the time warp associated with solution $s'$. The Parameter $\alpha^p$ is an adaptive penalization parameter that self-adjusts during the ALNS search process. During the ALNS, we initialize $\alpha^p$ to $\alpha^p_{min}$ and limit it to the interval $[\alpha^p_{min},\alpha^p_{max}]$. At the end of each iteration, $\alpha^p$ is updated based on the feasibility of the incumbent solution $s_t$. If the incumbent solution $s_t$ contains no time warp, $\alpha^p$ is set to $max\{\alpha^p/\rho,\alpha^p_{min}\}$ to encourage our ALNS to find infeasible solutions to reduce the probability of becoming trapped in a local optimum. Otherwise, if the incumbent solution is infeasible $\alpha$ is set to $min\{\alpha^p \cdot \rho,\alpha^p_{max}\}$ to guide the search to more feasible solutions. In this process, the parameter $\rho \geq 1$ controls to what extent $\alpha^p $ is adjusted. After every $\eta^{Reset}$ iteration, $\alpha^p$ is set back to $\alpha^p_{min}$ so that it does not get stuck at $\alpha^p_{max}$.

\subsection{Removal and Insertion Operators}\label{sec:removal_and_insertion}

\subsubsection{Removal Procedures}\label{sec:removal}
During each of the following removal procedures, we remove $\delta$ customer requests, where $\delta$ is drawn from the interval $[\omega_{min},\omega_{max}]\cdot min(|V_c|,100)$. The parameters $\omega_{min}$ and $\omega_{max}$ denote thereby the minimum and maximum share of customer requests to be removed. All removed customer requests are placed in the set of absent customer requests $B$. The following removal procedures can be used in our ALNS:

\begin{itemize}
\item[]\textbf{Random Removal}: a removal heuristic that randomly removes $\delta$ customer requests from a given solution $s'$ using a uniform probability distribution.

\item[]\textbf{Route Removal}: a removal heuristic where a random route is selected and up to $\delta$ customer requests of the route are randomly removed, using a uniform probability distribution. This is repeated until $\delta$ customers are removed in total \citep{Nagata.2009}.

\item[]\textbf{Worst Removal}: a removal heuristic introduced by \cite{Ropke.2006b} that removes customer requests with high costs that strongly contribute to the objective function of the current solution. To do so, we calculate for each customer request $cr_i$ the savings of removing it from $s'$ and sort the customer requests in descending order. Subsequently, we use a randomized removal controlled by a parameter $p_{worst}$, as proposed in \cite{Ropke.2006b} to remove the customer requests. This is done repeatedly until $\delta$ requests have been removed from $s'$.

\item[]\textbf{Historical Knowledge Node Removal}: removal heuristic that saves the lowest costs associated with each customer request and removes the $\delta$ customer requests with the highest difference between their current costs and saved historical lowest cost \citep{Demir.2012}. We adapted the procedure to include the transshipment costs of customer requests.

\item[]\textbf{Shaw Removal}: a removal heuristic that is also called related removal and was introduced by \cite{Shaw.1997}, \cite{Shaw.1998}. We define the similarity $S(i,j)$ between two customer requests $cr_i$ and $cr_j$ based on four customer request characteristics. First, their difference in demand $|q_{i}-q_{j}|$ divided by the maximum difference in demand $max_{i \in V_C}(q_i) - min_{i \in V_C}(q_i)$. Second, the distance $d_{ij}$ divided by the maximum distance between all customer requests $max_{i,j \in V_C}(d_{ij})$. Third, their absolute difference of time window centers $|tw_i^S+(tw_i^S-tw_i^E)/2-tw_j^S+(tw_j^S-tw_j^E)/2|$ divided by the maximum absolute difference between all time window centers. Fourth, the difference in the number of shared transshipment facilities $|F_i \cap F_j|$ divided by the maximum number of $max_{i,j \in V_C}(|F_i \cap F_j|)$. The last term, however, only makes a difference in instances where the possible transshipment facilities differ between customer requests. Each term is weighted by a weight factor $\chi_q,\chi_d,\chi_{TW}$ and $\chi_l$ respectively.

\item[]\textbf{Cluster Removal}:
introduced by \cite{Ropke.2006}, aims to remove an entire cluster of customer requests. Taking a given route, the customer requests in that route are partitioned into two clusters. Subsequently, one of these clusters is selected, and up to $\delta$ customer requests are removed. If the number of customer requests removed is smaller than $\delta$, a random customer request from the removed customer requests is selected. For this customer request, the current nearest customer request that is not in the same route is chosen. The route of the chosen customer request is again partitioned into two clusters and the procedure continues until $\delta$ customer requests have been removed from the solution.

\item[]\textbf{Distance-related Removal}: also called radial removal \citep{Schrimpf.2000}, is a special case of our Shaw Removal procedure, where only the distances between the currently assigned locations of the customer requests are used. The underlying idea of the removal procedure is to remove customer requests that are close to each other based on the distance between them. We implemented this procedure by randomly selecting a customer request and deleting it and its $\delta-1$-closest neighbors from the solution.

\item[]\textbf{Time-related Removal}: a removal heuristic that removes customer requests that are related in terms of the time at which they are served \citep{Pisinger.2007}.

\item[]\textbf{Adjacent String Removal}: introduced by \cite{Christiaens.2020} and showing promising results on classical vehicle routing benchmarks. String removal is based on the premise that removing only adjacent customers (e.g. as in distance-related removal) may lead to detours still being present in the destroyed route. Instead, removing adjacent strings of customer requests might be more efficient. During the procedure, either many strings with small cardinality or a few strings with a high cardinality can be removed. The upper limit of the string cardinality is thereby controlled by the parameter $L_{max}$. 
As described in \cite{Christiaens.2020}, the strings can be removed using either the \textit{string procedure} or \textit{split string procedure}. Contrary to the basic \textit{string procedure}, the \textit{split string procedure} preserves a random substring of customer requests from the string of customer requests to be removed. To control the number of customer requests to be preserved, the split depth parameter $\beta_{String}$ is used. The probability to select the \textit{split string procedure} instead of the \textit{string procedure} is controlled by the parameter $\alpha_{String}$.

\end{itemize}

\subsubsection{Insertion Procedures}\label{sec:insertion}
During the insertion phase of our algorithm, all locations per customer request are generally evaluated. To evaluate the insertions in constant time $\mathcal{O}(1)$, we store partial route information which depends on the type of problem considered. For our problem variants that do not include duration-based costs, we use the propositions from \cite{Schneider.2013} to evaluate the corresponding time window violations of the insertions. For the problem variants with duration-based costs and flexible vehicle departure times, we use the concatenation formulas described in Section \ref{sec:local_search} to determine the earliest and latest departures from the depot that lead to a schedule with minimum duration and time warp. Whenever a customer request is inserted into a route, the route information is updated in linear time, in relation to the number of customer requests in the route. This includes the time window violations up to and from each customer request within the route. The capacity usage is updated in constant time. Furthermore, in the case of heterogeneous fleets we use two ideas from \cite{Koc.2014}. First, before starting an insertion procedure, we check whether the destroyed routes can be served by smaller vehicles with lower fixed costs. Second, when inserting a customer request in a route would exceed the available vehicle capacity, we consider using a larger vehicle for the route and account for the difference in fixed costs.

In the following, we briefly describe the implemented insertion procedures:

\begin{itemize}

\item[]\textbf{Random Order Best Insertions}: all absent customer requests are sequentially inserted at their best insertion position in random order \citep{Christiaens.2020}.

\item[]\textbf{Demand Order Best Insertions}: inserts the absent customer requests sequentially at their best insertion position ordered by decreasing demand \citep{Christiaens.2020}.

\item[]\textbf{Farthest First Best Insertions}: inserts the absent customer requests sequentially at their best insertion position ordered by decreasing distance to the depot \citep{Christiaens.2020}.

\item[]\textbf{Closest First Best Insertions}: inserts the absent customer requests sequentially at their best insertion position ordered by increasing distance to the depot \citep{Christiaens.2020}.

\item[]\textbf{Regret-2 Insertion}: the regret-2 insertion was introduced by \citep{Ropke.2006} in the context of ALNS. To determine the insertion position of a customer request, a regret-2-value is calculated which is defined as the difference between its second-best and best insertion position. At each step of the insertion procedure, the regret-2-values are updated and the customer request with the highest difference is inserted at its best insertion position.
\end{itemize}

\subsubsection{Removal and Insertion Diversification}\label{diversification}
In order to randomize the removal and insertion procedures, we implement removal and insertion diversification strategies from the literature that can be used during the removal and insertion phases of our ALNS. First, we utilize the concept of adding a noise term to the objective function value during insertion \citep{Ropke.2006}. Each time an insertion position is evaluated, the insertion cost $\Delta C$ is modified with a probability $p_{noise}$. To do so, we draw a random factor $\xi$ from the interval $[\xi_{min},\xi_{max}]$ and calculate the modified insertion costs $\Delta C'= max\{0, \Delta C (1+\xi)\}$.

Second, inspired by the algorithm of \cite{Christiaens.2020}, we implement the concept of blinks as an insertion diversification strategy. Following \cite{Dumez.2021_trb}, we also extend the concept to the removal phase. The underlying idea of the concept is to skip possible removals and insertions with a defined probability. However, contrary to the insertion blinking procedure of \cite{Christiaens.2020}, where insertion positions within a tour can be skipped, we skip the evaluation of the possible location of a customer request when checking an insertion position within a route. Based on preliminary tests, we distinguish between the original customer location and the transshipment facilities and assign a blink probability $p_{Blinks-v_i}$ for the original customer locations and a blink probability $p_{Blinks-v_i}$ for the transshipment facilities.

for each a blink probability $p_{Blinks-v_i}$ and $p_{Blinks-V_F}$. Furthermore, we test if it is advantageous to avoid blinking on transshipment facilities when the previous or next customer request within the route is assigned to the transshipment facility to be evaluated. For the blinking of removal positions, we define $p_{Blinks-Removal}$ as the probability to skip a customer request during a removal procedure. To sum up, we implemented the following three blinking procedures for our problem: 
\begin{itemize}
\item[]\textbf{Blinks-R}: Blinking on customer requests to be removed.
\item[]\textbf{Blinks-I (I)}: Blinking for original customer locations.
\item[]\textbf{Blinks-I (II)}: Blinking for transshipment facilities.
\item[]\textbf{Blinks-I (III)}: Blinking for transshipment facilities unless they equal the previous or next customer's location within the route.
\end{itemize}

It should be noted that the methods not only serve for diversification but can also have a considerable effect on the computing time. In particular, increasing $p_{Blinks-V_F}$ can significantly reduce the computation time on problem instances with many possible transshipment facilities per customer request.

\subsection{Local Search Procedure}\label{sec:local_search}
In order to intensify the search, the approach described in this paper combines the ALNS general search methodology with a local search method. During each iteration of our ALNS, a local search procedure can be run with a probability $p_{LS}$ to improve the current solution $s'$. The local search during the ALNS is performed by a randomized variable neighborhood descent (RVND) \citep{Mladenovic.1997} which has been applied for many routing problems, especially in the context of heterogeneous fleets (see e.g. \cite{Penna.2019}, \cite{Subramanian.2010}).

The underlying idea of the RVND can be described as follows: Let $N = \{N^1,..., N^r\}$ be the set of neighborhood structures. Each time a selected neighborhood of the set of neighborhoods $N$ fails to improve the incumbent solution, the RVND randomly chooses another neighborhood from the same set to continue the search throughout the solution space. Algorithm \ref{alg:algorithm_RVND} gives an overview of the RVND procedure. During the first step of the algorithm, the list of inter-route neighborhoods $N_{Inter}$ is initialized (line 2). After the initial setup, for each iteration a random neighborhood $N^{(i)} \in N_{Inter}$ is selected and evaluated. Thereby, either a first- or best-improvement strategy can be utilized (line 5). If an improvement of the incumbent solution $s$ is found, $s$ is updated and an \textit{IntraRouteSearch} (here also a RVND) is initiated (line 6-8). Subsequently, the list of inter-route neighborhoods $N_{Inter}$ is refilled with all possible inter-route neighborhoods (line 9), and in the case of problems with varying fleet size and mix the available empty vehicles are updated (line 10). In contrast, if no improvement is found, $N^{(i)}$ is removed from $N_{Inter}$.

\begin{algorithm}[htb]
\begin{footnotesize}
	\caption{RVND}\label{alg:algorithm_RVND}
	\begin{algorithmic}[1]
	\Function{RVND}{$s$}
		\State Initialize the inter-route neighborhood List ($N_{Inter}$)
		\While{$N_{Inter} \neq \emptyset $}
			\State $N^{(i)} \gets$ Choose a random neighborhood $N^{(i)} \in N_{Inter}$
			\State $s' \gets$ Find the best- or first-improving neighbor $s'$ of $\in N^{(i)}$
			\If{$f(s')<f(s)$}
				\State $s\gets s'$	
				\State $s\gets$	IntraRouteSearch($s$)
				\State Update $N_{Inter}$
				\State Update Fleets \Comment only for fleet-size-and-mix instances 
			\Else 
			\State Remove $N^{(i)}$ from $N_{Inter}$
			\EndIf
			
		\EndWhile
	\EndFunction
	\end{algorithmic}
	\end{footnotesize}
\end{algorithm}

To evaluate the moves of the RVND efficiently, we use the concatenation strategies of \cite{Vidal.2014} and \cite{Vidal.2015}. These strategies are based on the idea that any route generated from a classical move applied on an incumbent solution $s'$ corresponds to a recombination of a bounded number of depot and customer request sequences of $s'$. As such, every new route can be expressed as a concatenation of sequences $\sigma_{1} \oplus \dots \oplus \sigma_{b}$. Each sequence $\sigma$ can be described by six values: distance $DIST(\sigma)$, demand $Q(\sigma)$, duration $D(\sigma)$, time warp $TW(\sigma)$, and the earliest visit $E(\sigma)$ and latest visit $L(\sigma)$ to the first vertex of $\sigma$ that lead to a schedule with minimum duration and time warp. Let $\sigma_{1}$ and $\sigma_{2}$ be two sequences. Using the equations of \cite{Vidal.2014} and \cite{Vidal.2015}, the concatenation of $\sigma_{1} \oplus \sigma_{2}$ can be formulated as:

\begin{flalign} \label{eq:concat}
DIST(\sigma_1 \oplus \sigma_2) &= DIST(\sigma_1) + d\sigma_1(|\sigma_1|)\sigma_2(1) + DIST(\sigma_2)\\
Q(\sigma_1 \oplus \sigma_2) &= Q(\sigma_1) + Q(\sigma_2)\\
D(\sigma_1 \oplus \sigma_2) &= D(\sigma_1) + t\sigma_1(|\sigma_1|)\sigma_2(1) + D(\sigma_2) + \Delta WT\\
E(\sigma_1 \oplus \sigma_2) &= max\{E(\sigma_2) -\Delta, E(\sigma_1)\} - \Delta WT\\
L(\sigma_1 \oplus \sigma_2) &= min\{L(\sigma_2) -\Delta, L(\sigma_1))\} + \Delta TW\\
TW(\sigma_1 \oplus \sigma_2) &= TW(\sigma_1) + TW(\sigma_2) + \Delta TW\\
where ~ \Delta &= D(\sigma_1) - TW(\sigma_1) + t\sigma_1(|\sigma_1|)\sigma_2(1)\\
\Delta WT &= max\{E(\sigma_2) - \Delta - L(\sigma_1), 0\}\\
\Delta TW &= max\{E(\sigma_1) + \Delta - L(\sigma_2), 0\}.
\end{flalign}

\subsubsection{Inter- and Intra-Route Neighborhood Structures}\label{sec:ls_neighborhoods}

The definition of neighborhood structures is one of the main aspects in the design of local search algorithms. In order to achieve better local optima, we use both inter-route and intra-route neighborhoods in our search. To reduce the computational effort of our inter-route swap neighborhoods, we apply the pruning mechanism for vehicle routing problems with time windows from \cite{Vidal.2013}. The correlation between two locations is, thereby, calculated by the weighted sum of the distance, minimum waiting time, and minimum penalty between the two locations, using the same weights as in \cite{Vidal.2013}. Subsequently, for each location $l \in V'$ we store the $\epsilon=30$ highest-correlated locations in an immutable neighbor list. Furthermore, for each move, the capacity feasibility is evaluated first and the move is discarded if it is infeasible.

We employ the following seven inter-route neighborhoods in our local search procedure:

\begin{itemize}
\item[•]\textbf{Shift(1,0)} transfers a customer request from one route to another.
\item[•]\textbf{Shift(2,0)} transfers two adjacent customer requests from one route to another.
\item[•]\textbf{2-Opt*} inter-route version of the classical 2-Opt move \citep{Potvin.1995}.
\item[•]\textbf{Swap(1,1)} permutation between two customer requests from different routes.
\item[•]\textbf{Swap(2,1)} permutation of two adjacent customer requests from one route by one customer request from another route.
\item[•]\textbf{Swap(2,2)} permutation of two adjacent customer requests from one route by two customer requests from another route.
\item[•]\textbf{$K$-Shift} transfers $K$ adjacent customer requests from one route to the end of another route or a new route of a vehicle with lower fixed costs \citep{Penna.2013}. In the case of unlimited fleets, we ensure that there is at least one empty vehicle of each type. 
\end{itemize}

Regarding the intra-route neighborhoods, we implement the well-known \textbf{2-Opt} neighborhood \citep{Lin.1965}, \textbf{Reinsertion}, \textbf{Or-opt-2}, and \textbf{Or-opt-3} \citep{Or.1976} neighborhoods, as well as an intra-route version of \textbf{Swap(1,1)}. In addition to these well-known general neighborhoods, we implement new specific neighborhoods for vehicle routing problems with transshipment facilities. These intra-route neighborhoods aim to modify the assignment of transshipment facilities to customer requests within a route. The underlying idea of these operators is similar to the delivery option moves of \cite{Zhou.2018}, where the assignment of customer requests to intermediate pickup facilities is modified with three move types. However, unlike \cite{Zhou.2018}, where the current position of customers within the routes is not considered, and either a single customer request or the requests of the n-closest customers to a transshipment facility are transshipped, our intra-route neighborhoods focus on the modification of the locations of successive visits within a single route.

\begin{figure}[htb]
\centering
\begin{tikzpicture}[x=0.40mm, y=0.40mm, inner xsep=0pt, inner ysep=0pt, outer xsep=0pt, outer ysep=0pt]
\path[line width=0mm] (-24.68,32.14) rectangle +(218.57,75.40);
\definecolor{L}{rgb}{0,0,0}
\definecolor{F}{rgb}{0.753,0.753,0.753}
\path[line width=0.30mm, draw=L, fill=F] (40.00,100.00) [rotate around={270:(40.00,100.00)}] rectangle +(20.00,100.00);
\definecolor{F}{rgb}{0.878,0.878,0.878}
\path[fill=F] (4.68,100.09) -- (34.68,100.09) -- (34.68,80.09) -- (4.68,80.09) -- cycle;
\definecolor{L}{rgb}{1,1,1}
\path[line width=0.15mm, draw=L] (4.68,96.60) -- (8.17,100.09) (4.68,89.53) -- (15.24,100.09) (4.68,82.46) -- (22.31,100.09) (9.38,80.09) -- (29.38,100.09) (16.45,80.09) -- (34.68,98.32) (23.53,80.09) -- (34.68,91.25) (30.60,80.09) -- (34.68,84.18);
\definecolor{L}{rgb}{0,0,0}
\path[line width=0.30mm, draw=L] (4.68,100.09) -- (34.68,100.09) -- (34.68,80.09) -- (4.68,80.09);
\path[fill=F] (173.97,80.09) -- (143.97,80.09) -- (143.97,100.09) -- (173.97,100.09) -- cycle;
\definecolor{L}{rgb}{1,1,1}
\path[line width=0.15mm, draw=L] (143.97,94.48) -- (149.59,100.09) (143.97,87.41) -- (156.66,100.09) (143.97,80.33) -- (163.73,100.09) (150.80,80.09) -- (170.80,100.09) (157.88,80.09) -- (173.97,96.19) (164.95,80.09) -- (173.97,89.12) (172.02,80.09) -- (173.97,82.05);
\definecolor{L}{rgb}{0,0,0}
\path[line width=0.30mm, draw=L] (173.97,80.09) -- (143.97,80.09) -- (143.97,100.09) -- (173.97,100.09);
\definecolor{F}{rgb}{0.753,0.753,0.753}
\path[fill=F] (-8.47,102.54) arc (0:0:0.00mm);
\definecolor{L}{rgb}{1,1,1}
\path[line width=0.15mm, draw=L] ;
\definecolor{L}{rgb}{0,0,0}
\path[line width=0.30mm, draw=L] (-8.47,102.54) arc (0:0:0.00mm);
\path[line width=0.30mm, draw=L] (5.00,90.00) -- (174.00,90.00);
\path[line width=0.30mm, draw=L] (5.13,90.00) .. controls (1.12,90.59) and (-2.63,92.32) .. (-5.68,95.00) .. controls (-7.31,96.44) and (-8.71,98.13) .. (-9.82,100.00);
\path[line width=0.30mm, draw=L] (173.94,90.01) .. controls (177.95,90.60) and (181.71,92.33) .. (184.75,95.01) .. controls (186.39,96.45) and (187.79,98.14) .. (188.90,100.01);
\definecolor{F}{rgb}{0,0,0}
\path[line width=0.30mm, draw=L, fill=F] (188.90,100.01) -- (186.87,97.96) -- (188.07,97.25) -- (188.90,100.01) -- cycle;
\definecolor{F}{rgb}{1,1,1}
\path[line width=0.30mm, draw=L, fill=F] (50.01,89.65) circle (2.87mm);
\path[line width=0.30mm, draw=L, fill=F] (90.19,89.83) circle (2.87mm);
\path[line width=0.30mm, draw=L, fill=F] (70.00,90.00) circle (2.83mm);
\path[line width=0.30mm, draw=L, fill=F] (110.00,90.00) circle (2.83mm);
\path[line width=0.30mm, draw=L, fill=F] (130.00,90.00) circle (2.83mm);
\path[line width=0.30mm, draw=L, fill=F] (156.49,90.00) circle (2.83mm);
\path[line width=0.30mm, draw=L, fill=F] (20.80,90.00) circle (2.83mm);
\definecolor{F}{rgb}{0.753,0.753,0.753}
\path[fill=F] (-8.47,62.88) arc (0:0:0.00mm);
\definecolor{L}{rgb}{1,1,1}
\path[line width=0.15mm, draw=L] ;
\definecolor{L}{rgb}{0,0,0}
\path[line width=0.30mm, draw=L] (-8.47,62.88) arc (0:0:0.00mm);
\draw(20.86,91.03) node[anchor=base]{\fontsize{5}{5.46}\selectfont $cr_{13}$};
\draw(20.86,85.94) node[anchor=base]{\fontsize{5}{5.46}\selectfont $v_{13}$};
\draw(49.84,91.03) node[anchor=base]{\fontsize{5}{5.46}\selectfont $cr_6$};
\draw(49.84,85.94) node[anchor=base]{\fontsize{5}{5.46}\selectfont $v_6$};
\draw(69.70,91.03) node[anchor=base]{\fontsize{5}{5.46}\selectfont $cr_3$};
\draw(69.70,85.94) node[anchor=base]{\fontsize{5}{5.46}\selectfont $v_3$};
\draw(90.03,91.03) node[anchor=base]{\fontsize{5}{5.46}\selectfont $cr_4$};
\draw(90.03,85.94) node[anchor=base]{\fontsize{5}{5.46}\selectfont $v_4$};
\draw(110.13,91.03) node[anchor=base]{\fontsize{5}{5.46}\selectfont $cr_{11}$};
\draw(110.13,85.94) node[anchor=base]{\fontsize{5}{5.46}\selectfont $v_{11}$};
\draw(130.00,91.03) node[anchor=base]{\fontsize{5}{5.46}\selectfont $cr_2$};
\draw(130.00,85.94) node[anchor=base]{\fontsize{5}{5.46}\selectfont $v_2$};
\draw(156.52,91.03) node[anchor=base]{\fontsize{5}{5.46}\selectfont $cr_8$};
\draw(156.52,85.94) node[anchor=base]{\fontsize{5}{5.46}\selectfont $v_i$};
\path[line width=0.30mm, draw=L, fill=F] (40.00,57.14) [rotate around={270:(40.00,57.14)}] rectangle +(20.00,100.00);
\definecolor{F}{rgb}{0.878,0.878,0.878}
\path[fill=F] (4.68,57.23) -- (34.68,57.23) -- (34.68,37.23) -- (4.68,37.23) -- cycle;
\definecolor{L}{rgb}{1,1,1}
\path[line width=0.15mm, draw=L] (4.68,54.18) -- (7.74,57.23) (4.68,47.10) -- (14.81,57.23) (4.68,40.03) -- (21.88,57.23) (8.95,37.23) -- (28.95,57.23) (16.02,37.23) -- (34.68,55.89) (23.09,37.23) -- (34.68,48.82) (30.16,37.23) -- (34.68,41.75);
\definecolor{L}{rgb}{0,0,0}
\path[line width=0.30mm, draw=L] (4.68,57.23) -- (34.68,57.23) -- (34.68,37.23) -- (4.68,37.23);
\path[fill=F] (173.97,37.23) -- (143.97,37.23) -- (143.97,57.23) -- (173.97,57.23) -- cycle;
\definecolor{L}{rgb}{1,1,1}
\path[line width=0.15mm, draw=L] (143.97,52.05) -- (149.16,57.23) (143.97,44.98) -- (156.23,57.23) (143.97,37.91) -- (163.30,57.23) (150.37,37.23) -- (170.37,57.23) (157.44,37.23) -- (173.97,53.77) (164.51,37.23) -- (173.97,46.70) (171.58,37.23) -- (173.97,39.62);
\definecolor{L}{rgb}{0,0,0}
\path[line width=0.30mm, draw=L] (173.97,37.23) -- (143.97,37.23) -- (143.97,57.23) -- (173.97,57.23);
\definecolor{F}{rgb}{0.753,0.753,0.753}
\path[fill=F] (-8.47,59.68) arc (0:0:0.00mm);
\definecolor{L}{rgb}{1,1,1}
\path[line width=0.15mm, draw=L] ;
\definecolor{L}{rgb}{0,0,0}
\path[line width=0.30mm, draw=L] (-8.47,59.68) arc (0:0:0.00mm);
\path[line width=0.30mm, draw=L] (5.00,47.14) -- (174.00,47.14);
\path[line width=0.30mm, draw=L] (5.13,47.14) .. controls (1.12,47.72) and (-2.63,49.46) .. (-5.68,52.14) .. controls (-7.31,53.58) and (-8.71,55.27) .. (-9.82,57.14);
\path[line width=0.30mm, draw=L] (173.94,47.15) .. controls (177.95,47.74) and (181.71,49.47) .. (184.75,52.15) .. controls (186.39,53.59) and (187.79,55.28) .. (188.90,57.15);
\definecolor{F}{rgb}{0,0,0}
\path[line width=0.30mm, draw=L, fill=F] (188.90,57.15) -- (186.87,55.10) -- (188.07,54.39) -- (188.90,57.15) -- cycle;
\definecolor{F}{rgb}{1,1,1}
\path[line width=0.30mm, draw=L, fill=F] (50.01,46.79) circle (2.87mm);
\path[line width=0.30mm, draw=L, fill=F] (90.19,46.97) circle (2.87mm);
\path[line width=0.30mm, draw=L, fill=F] (70.00,47.14) circle (2.83mm);
\path[line width=0.30mm, draw=L, fill=F] (110.00,47.14) circle (2.83mm);
\path[line width=0.30mm, draw=L, fill=F] (130.00,47.14) circle (2.83mm);
\path[line width=0.30mm, draw=L, fill=F] (156.49,47.14) circle (2.83mm);
\path[line width=0.30mm, draw=L, fill=F] (20.80,47.14) circle (2.83mm);
\draw(20.86,48.17) node[anchor=base]{\fontsize{5}{5.46}\selectfont $cr_{13}$};
\draw(20.86,43.19) node[anchor=base]{\fontsize{5}{5.46}\selectfont $v_{13}$};
\draw(49.84,48.17) node[anchor=base]{\fontsize{5}{5.46}\selectfont $cr_6$};
\draw(49.84,43.19) node[anchor=base]{\fontsize{5}{5.46}\selectfont $v_f$};
\draw(69.70,48.17) node[anchor=base]{\fontsize{5}{5.46}\selectfont $cr_3$};
\draw(69.70,43.19) node[anchor=base]{\fontsize{5}{5.46}\selectfont $v_f$};
\draw(90.03,48.17) node[anchor=base]{\fontsize{5}{5.46}\selectfont $cr_4$};
\draw(90.03,43.19) node[anchor=base]{\fontsize{5}{5.46}\selectfont $v_f$};
\draw(110.13,48.17) node[anchor=base]{\fontsize{5}{5.46}\selectfont $cr_{11}$};
\draw(110.13,43.19) node[anchor=base]{\fontsize{5}{5.46}\selectfont $v_f$};
\draw(130.00,48.17) node[anchor=base]{\fontsize{5}{5.46}\selectfont $cr_2$};
\draw(130.00,43.19) node[anchor=base]{\fontsize{5}{5.46}\selectfont $v_f$};
\draw(156.52,48.17) node[anchor=base]{\fontsize{5}{5.46}\selectfont $cr_8$};
\draw(156.52,43.19) node[anchor=base]{\fontsize{5}{5.46}\selectfont $v_8$};
\draw(-18.36,48.38) node[anchor=base]{\fontsize{7}{8.19}\selectfont $r'$};
\draw(-18.68,86.74) node[anchor=base]{\fontsize{7}{8.19}\selectfont $r$};
\path[line width=0.30mm, draw=L] (90.00,75.00) -- (90.00,62.00);
\definecolor{F}{rgb}{0,0,0}
\path[line width=0.30mm, draw=L, fill=F] (90.00,75.00) -- (89.30,72.20) -- (90.70,72.20) -- (90.00,75.00) -- cycle;
\path[line width=0.30mm, draw=L, fill=F] (90.00,62.00) -- (90.70,64.80) -- (89.30,64.80) -- (90.00,62.00) -- cycle;
\end{tikzpicture}
\caption{Example of the two moves for changing the location of larger parts of routes. The customer requests of the sequence $ cr_6\dots cr_2$ share at least one possible transshipment facility $v_f \in V_F$ that is neither shared by the upstream nor downstream customer requests $cr_{13}$ and $cr_{8}$ of the tour ($v_f \notin F_{13} \cup F_8$).}\label{fig:changeLocationMove}
\end{figure}
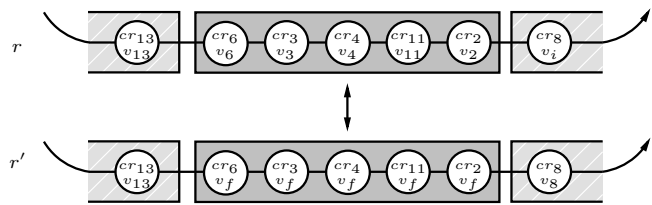

The first three neighborhoods, named \textbf{ChangeLocation($m$)}, aim for small changes and are based on the idea of changing the location of $m$ consecutive customer requests, which share at least one common transshipment facility in their set of possible transshipment facilities $F$, to one of the common transshipment facilities or to the respective customer locations. We thereby implemented versions for $m=$ $1$, $2$, and $3$ customer requests to be changed simultaneously. Due to our observation that the decision for or against transshipping goods at a transshipment facility often depends on a critical number of customer requests per tour being assigned to the same transshipment facility, we implemented two additional moves to change the locations of more customer requests within a route $r$. This modification of customer locations is an important mechanism, as depending on the problem instance, the solution can become stuck in local optima, where all or most customer requests of a tour are allocated to the same transshipment facility and the previous local search moves are not large enough to leave the local optima. Likewise, it is also possible that using a transshipment facility is only cost-effective if a large portion or all customer requests in an area are assigned to it. Thus, the two moves aim to change the assigned locations of longer sequences of suitable customer requests within a tour (see Figure~\ref{fig:changeLocationMove}). The first neighborhood, \textbf{Undo-Transshipment}, filters for sequences of adjacent customer requests being assigned to the same transshipment facility and assigns each customer request to its respective customer location. The second neighborhood, \textbf{Try-Transshipment}, is its counterpart and identifies sequences of adjacent customer requests that share a common transshipment facility $v_f \in V_F$ and assigns the customer request to the facility. In the case of problem instances where each customer request can be assigned to every transshipment facility, the sequence of customer requests changed equals the entire route. As a consequence, it is possible to limit the maximum sequence length for routes that include many customers (e.g. maximum 10 customer requests) and evaluate every substring.

With the aim to explore the location assignment of customer requests more thoroughly during the local search, we also propose \textit{combined neighborhoods} that jointly change the assigned location as well as the visit sequence of customer requests. We thereby consider combined versions of the \textbf{Shift(1,0)} \textbf{Reinsertion}, and inter-route and intra-route \textbf{Swap(1,1)} neighborhoods. The combined \textbf{Reinsertion} neighborhood is very similar to the \textit{remove a CP service} neighborhood of \cite{Zhou.2018}, with the exception that in the latter only customers which are currently assigned to a transshipment facility are considered. Whereas our preliminary results have shown that the addition of combined neighborhoods can lead to slight improvements in solution quality in some cases, it also significantly increases the computation times. This applies especially to instances where the customer requests have many possible transshipment locations, such as the VRPTF benchmark instances from \cite{Baldacci.2017}. For this reason, combined neighborhoods are not considered further in this paper.

\subsection{Set-Partitioning Problem}\label{sec:spp}

Every $\eta^{SP}$ iteration, the pool of stored routes $R^{pool}$ is used to build a restricted set-partitioning (SP) model to be solved by a mixed integer programming (MIP) solver (see Equations (\ref{eq:scp_1})-(\ref{eq:scp_4})). Based on the notation of \cite{Subramanian.2012}, $R^{pool}_i \subseteq R^{pool}$ denotes the subset of routes that contain customer request $cr_i$ and $R^{pool}_k \subseteq R^{pool}$ the subset of routes that use vehicle type $m_k$. The binary variable $z_r$ indicates whether a route is included in the solution or not.

\begin{flalign} \label{eq:scp_1}
min \quad & \sum_{r \in R^{pool}} c_r z_r\\
\label{eq:scp_2} s. t. \quad & \sum_{r \in R^{pool}_i} z_r =1 \quad \forall cr_i \in CR\\ 
 \label{eq:scp_3} &\sum_{r \in R^{pool}_k} z_r \leq m_k \quad \forall cr_i \in CR\\
&z_r \in \{0,1\} \quad \forall r \in R^{pool}. \label{eq:scp_4}
\end{flalign}

The objective of the SP model (\ref{eq:scp_1}) is to minimize the sum of the route costs. Constraint (\ref{eq:scp_2}) ensures that each customer request is contained exactly once in the solution. Constraint (\ref{eq:scp_3}) states the constraint for the number of available vehicles per type. However, for the fleet size and mix problem variants, constraint (\ref{eq:scp_3}) is left out because the number of vehicles per type is not restricted. Furthermore, as \cite{Dumez.2021} point out, the set-partitioning constraint (\ref{eq:scp_2}) can be modified to an easier to solve set-covering constraint ($\geq1$ instead of $=1$) if the triangle inequality holds true for the problem instance that is considered. If the resulting solution of the set-covering problem contains a customer request in more than one route, we remove the redundant occurrences based on a simple greedy procedure. 

To reduce the computational effort and time, the solver is initialized with $S_{best}$, and the maximum computation time is limited to 30 seconds.

\subsection{Solution Acceptance}\label{sec:acceptance}
Solution acceptance criteria decide whether a new solution $s'$ is accepted to replace the incumbent solution $s_t$. There is a broad variety of acceptance criteria that have been tested with the ALNS framework \citep{Santini.2018}. In the following, we describe two implemented acceptance criteria in our ALNS:

\begin{itemize}
\item[]\textbf{Threshold Acceptance}: Threshold acceptance (TA) was introduced by \cite{Dueck.1990}. As the name indicates, each new solution $s'$ is accepted if the difference in solution quality $f(s')-f(s_t)/f(s')$ is smaller than a defined threshold $\zeta$. With each iteration, the threshold is reduced until it reaches an end value. To determine the decrease in each iteration, either linear or exponential schemes, as shown in \cite{Schrimpf.2000}, can be used. The linear threshold is thereby calculated as $\zeta= \zeta_{0}(1-\frac{k}{\eta^{max}})$, where $\zeta_{0}$ is the initial threshold and $\frac{k}{\eta^{max}}$ the percentage of maximum iterations finished. The exponential threshold is calculated as $\zeta= \zeta_{0}~exp(-ln(2) \frac{k}{\eta^{max}ta_{\alpha}})$, with $ta_{\alpha}$ being a factor controlling the half-lives. The initial threshold $\zeta_{0}$ is a parameter that has to be specified.

\item[]\textbf{Record-to-Record Acceptance}: Record-to-record acceptance (RRA) was introduced by \cite{Dueck.1993} and is based on a similar idea to TA. However, instead of comparing the current solution $s'$ against the incumbent solution $s_t$, RRA compares the difference in solution values between the current solution $s'$ and best solution $s_{best}$. If $(f(s')-f(s_{best}))/f(s')$ is smaller than the current threshold, the solution is accepted. As with TA, the threshold can be decreased linearly or exponentially.
\end{itemize}

\section{Computational Results}\label{sec:results}
In this section, we present the computational results for our ALNS and analyze its performance compared to other approaches in the literature. We coded our ALNS as a single-threaded code in Java 11. To solve the SPP, we used IBM Ilog CPLEX 12.10.0 as the MIP solver. All experiments were performed on an Intel Core i5-6200U CPU at 2.3 GHz with 8 GB RAM on a Windows 10 operating system.

Section~\ref{sec:param_selection} presents the process of selecting the parameters of the ALNS, including the decision of which algorithmic components to use. Following this, in Section~\ref{sec:instances_literature} we derive and describe benchmark instances from the literature and compare the results where applicable. Finally, in Section~\ref{sec:real-world_instances} we introduce a real-world instance and analyze the impact of transshipment fees, time windows, heterogeneous fleets, and demand size on the transshipment decisions.

\subsection{Parameter Selection}\label{sec:param_selection}

To tune the parameters of our algorithm, we used the automatic algorithm configuration tool irace (version 3.3) \citep{LopezIbanez.2016}. The irace tool implements an iterated racing procedure that provides a set of parameter configurations, so-called elite configurations, which statistically prove to be suitable. To configure our ALNS with irace, the parameters to be tuned, and their types and value ranges must be defined. Additionally, a set of training instances and the tuning budget (total number of runs to be performed) must be specified. We set the tuning budget for our algorithm to \num{20000} runs with \num{5000} to \num{40000} maximum iterations per run, depending on the problem. Each run corresponds to the execution of one sampled parameter configuration, given a random seed and a randomly selected training instance. For the training instances, we select instances of varying sizes, including instances with homogeneous and heterogeneous fleets, as well as time windows. Due to the heterogeneity of our training instances, we set irace to use an $F$-test with a confidence level of 0.95 as the statistical test to analyze the differences between the algorithm configurations. Table~\ref{tab:irace_input} shows the parameter types as well as their possible ranges and best-found values in irace. Thereby, we distinguish between the VRPTF and its extensions with time windows. Furthermore, we not only determined the parameters of our ALNS but also evaluated the suitability of our two acceptance criteria (see Section~\ref{sec:acceptance}). In addition, we also let irace decide which removal and insertion diversification procedures to include in our ALNS. This is in line with the studies of \cite{Francois.2019} and \cite{Francois.2016}, who argue that an a priori selection of algorithmic components is important when designing a ALNS.

\begin{table}[h]
\footnotesize
\centering
\caption{Parameters in the ALNS configuration.\label{tab:irace_input}}
{\begin{tabular*}{\textwidth}{ @{\extracolsep{\fill}}llllrr}\toprule
Category & Parameter & Type & Authorized Range & $\neg$ TW & TW\\
\midrule
\multirow{ 7}{*}{General ALNS} &$\eta^{update}$ & Integer, Step 100 &[100, \num{10000}] & 600 & 3500\\
&$\lambda$ &Rea1 & [0.1, 1]& 0.49 & 0.73\\
&$\alpha^p_{min}$ &Integer, Step 5 & $[5,500]$& -- &30\\
&$\alpha^p_{max}$ &Integer, Step 10 & [$\alpha^p_{min}$, \num{1000}]& -- &900\\
&$\rho$ &Real, Step 0.1 & [1, 2]& -- &1.5\\
&$\eta^{Reset}$ &Integer, Step 100 & [100, \num{5000}]& -- & 800\\
&$\eta^{SP}$ &Integer, Step 5000 & [5000, \num{20000}]& \num{5000} & \num{15000}\\
\midrule
\multirow{ 11}{*}{Removal} & $\omega_{min}$ & Rea1 & [0.05,0.25]&0.11 & 0.07\\
& $\omega_{max}$ & Rea1 & [0.25, 0.6]& 0.42 & 0.31\\
& $p_{worst}$ & Integer & $[1,6]$ & 3  & 3\\
& $L_{max}$ & Integer & $[2,25]$ & 19 & 15\\
& $\alpha_{String}$ & Real, Step 0.1 & [0.2, 0.8] & 0.40 &0.45\\
& $\beta_{String}$ & Real & [0.01, 0.1] & 0.03 & 0.03\\
& $\chi_q$ & Real & [0, 1] & 0.42 & 0.39\\
& $\chi_d$ & Real & [0, 1] & 0.25 & 0.11\\
& $\chi_l$ & Real & [0, 1] & 0.33 &0.17\\
& $\chi_{TW}$ & Real & [0, 1] & --&0.33\\
& $p_{Blinks-Removal}$ & Rea1 & [0, 0.5]& 0.12 & 0.24\\
\midrule
\multirow{4}{*}{Insertion}& $p_{noise}$ & Rea1 & [0, 0.3] & 0.05 & 0.00\\
& $\xi_{max}$ & Rea1 & [0, 0.3]& 0.15 & 0.00\\
& $p_{Blinks-v_i}$ & Rea1 & [0, 0.2]& 0.02 & 0.01\\ 
& $p_{Blinks-V_F}$ & Rea1 & [0, 0.9]& 0.10 & 0.13\\ 
\midrule
\multirow{ 4}{*}{Acceptance}& Criterion & Categorical & $\{$TA, RRA$\}$& TA & RRA\\
&Threshold-Type & Categorical & $\{$linear, exponential$\}$& exponential & linear\\
&$ta_{\alpha}$& Rea1 & [0.05, 0.3]&0.12 & --\\
&$\zeta_{0}$& Rea1 & [0.01, 0.1]&0.02 &0.07\\
\bottomrule
\end{tabular*}}
\end{table}

The results of the final configurations in Table~\ref{tab:irace_selection} indicate that most of the implemented removal procedures, except the route removal procedure, are beneficial for the ALNS. However, some procedures, such as the cluster and time-related removal, were selected only in the configuration for instances with time windows. Looking at the diversification mechanisms from Section \ref{diversification}, it can be observed that blinking during the ruin step (Blinks-R) seems to be advantageous for instances with time windows as well as instances without time windows. The addition of noise, however, is only selected in the configuration for VRPTF instances. Blinking on customer requests locations (Blinks (I) and Blinks (II)) is chosen in both configurations. However, the variant Blinks (III) for insertion does not appear in any elite configuration for instances with time windows. Regarding the acceptance criteria, it can be seen that TA performed better on instances without time windows, whereas RRA did better on instances with time windows.

\begin{table}[h]
\footnotesize
\centering
\caption{ALNS Configurations.\label{tab:irace_selection}}
{\begin{tabular}{ llcc}\toprule  \multirow{2}{3cm}{Category} &\multirow{2}{3cm}{Component}& \multicolumn{2}{c}{Configurations}\\
\cline{3-4}\rule{0pt}{2.5ex}  
& & $\neg$ TW & TW\\
\midrule
\multirow{ 8}{*}{Removal Procedures} & Random Removal &\checkmark & \checkmark \\
&Route Removal & &  \\
&Worst Removal &\checkmark & \checkmark \\
&Shaw Removal &\checkmark & \checkmark \\
&Cluster Removal & &  \checkmark \\
&Historical Knowledge Node Removal &  \checkmark &  \checkmark \\
&Distance-related Removal &\checkmark & \checkmark  \\
&Time-related Removal& & \checkmark  \\
&Adjacent String Removal &\checkmark & \checkmark \\
\midrule
\multirow{ 5}{*}{Insertion Procedures} & Random Order &\checkmark &\checkmark\\
&Demand Order & \checkmark &  \checkmark  \\
&Farthest Order &   &  \checkmark \\
&Closest Order &   &  \checkmark \\
&Regret-2 & \checkmark &  \checkmark \\
\midrule
\multirow{5}{*}{Diversification} & Noise &\checkmark &  \\
&Blinks (I)&  \checkmark & \checkmark\\
&Blinks (II) & \checkmark &\checkmark\\
&Blinks (III)& \checkmark &  \\
&Blinks-R& \checkmark & \checkmark \\
\midrule
\multirow{ 2}{*}{Acceptance criteria} &TA& \checkmark\\
&RRA& &  \checkmark \\
\bottomrule
\end{tabular}}
\end{table}

In contrast to the previous parameters tuned by irace, the probability $p_{LS}$ of the embedded local search was selected manually using a maximum CPU time per Instance. In this connection, on the one hand, the possible benefits of high selection probabilities and on the other hand, the additional computing effort should be weighed up. Our tests showed that values of 0.125 resulted in promising results for $p_{LS}$, while larger values did not improve the solution quality within the same computation time.

\subsection{Problem Instances from the Literature}\label{sec:instances_literature}
To compare our metaheuristic with existing approaches from the literature, we consider several instances for the VRPTF and similar problems, such as the VRPRDL and VRPHRDL, from the literature. As there are no instances available for variants of the VRPTF with time windows and heterogeneous fleets, we propose new instances based on classical instances for the \textit{vehicle routing problem with time windows} (VRPTW) from the literature. 

The instances for the VRPTF and its variants vary by the number of customers $V_C$ and the number of transshipment facilities $V_F$. For each of these instances, every customer request can be assigned to every transshipment facility (i.e. $F_i=V_F$, $ \forall i \in V_C$). The cost of assigning a customer request to a transshipment facility is thereby calculated as the Euclidean distance $d_{if}$ between the customer's location $v_i \in V_C$ and the selected transshipment facility $v_f \in V_F$. For each instance, our ALNS is run with ten different random seeds until a maximum number of iterations is reached.

\subsubsection{VRPTF}\label{sec:instances_vrptf} 
In this section, we give an overview of our results on benchmark instances for the VRPTF and compare these to the results of \cite{Baldacci.2017}. Thereby, we focus on three sets of problem instances for the VRPTF that have been modified by \cite{Baldacci.2017} from classical instances for the \textit{location routing problem} (LRP). These 65 instances vary between 12 to 150 customer requests and feature between 2 to 20 transshipment facilities. Instance set (I) has been derived from \cite{Akca.2009} and includes 12 instances with either 30 or 40 customer requests and 12 transshipment facilities. Instance set (II) is based on the instances of \cite{Prins.2004} and features instances with 20 to 100 customer requests and either 5 or 10 transshipment facilities. Finally, instance set (III) involves various instances from LRP literature with instances varying between 12 to 150 customer requests and 2 to 20 transshipment facilities. As in \cite{Baldacci.2017}, the distances between locations are computed as Euclidean distances following the TSPLIB EUC\_2D standard.

For our analysis, the number of iterations per instance is set to $\eta^{max}=5000$ iterations. However, for many smaller instances, optimal solutions can be found in less than \num{1000} iterations. Table~\ref{tab:comparison_baldacci} summarizes the results obtained on the three sets and compares them to the results of \cite{Baldacci.2017}. In the table, column \#Opt. reports the number of instances on which the optimal solutions, as reported in \cite{Baldacci.2017}, were found. Column \#Imp. reports the number of instances per set where improvement was found and column \#Eq. states the number of instances where the same non-optimally proven results were obtained. Column Avg. Imp. states the average improvement in \%. Lastly, column Avg. $T$ Red. \% reports the average time reduction, based on the average run time from 10 runs each. The detailed results on each instance, including the best solution values $z*$, average solution values $\overline{z}$, and times $\overline{\text{T}}$ out of 10 runs can be found in \ref{sec:details_vrptf}.

Considering the reported solutions from \cite{Baldacci.2017} and their stated lower bounds, it can be seen that our ALNS heuristic can find new best solutions for 38 problem instances while only taking a fraction of the computational times. For the remaining 27 problem instances, solutions identical to those reported by \cite{Baldacci.2017} have been found. Of these, 23 have been proven to be optimal by \cite{Baldacci.2017}.

\begin{table*}[!htp]
\footnotesize
\center
\caption{Summary of the tested data sets from \cite{Baldacci.2017}.\label{tab:comparison_baldacci}}
{\begin{tabular}{ lc c c ccc c c}\toprule
Instance set & \#& \#Opt. & \#Imp. &\#Eq. & Avg. Imp. \% & Avg. T Red. \%\\
\midrule
(I) \cite{Akca.2009} & 12 & 10 & 0 & 2 & 0.00 & 85.16\\
(II) \cite{Prins.2004}& 14 & 3& 9&2 & 0.81 & 94.73\\
(III) Different authors & 39 & 10 & 29& 0 & 0.85 &94.66\\
\bottomrule
\end{tabular}}
\end{table*}

\subsubsection{VRPRDL and VRPHRDL}
In order to validate our algorithm's performance on problem instances with time windows, we utilize the 120 instances from the VRPRDL and VRPHRDL of \cite{Reyes.2017} and \cite{Ozbaygin.2017} and compare our algorithm with recent studies. To be precise, we compare the results of our ALNS on the instances to those reported by \cite{Ozbaygin.2017}, \cite{Dumez.2021_trb}, and \cite{Tilk.2020}. To compare our results to those reported by \cite{Ozbaygin.2017}, we set the maximum number of vehicles to the number reported by \cite{Ozbaygin.2017}, as done in \cite{Dumez.2021_trb}, and \cite{Tilk.2020}. In summary, as shown in Tables~\ref{tab:comparison_VRPRDL} and \ref{tab:comparison_VRPHRDL}, we can observe that the best-known values, as reported in the literature, are found for all 120 instances. Furthermore, with respect to the average solution quality and computing time $\overline{\text{T}}$, only very small differences to the current state-of-the-art algorithm from \cite{Dumez.2021_trb} can be observed.

\begin{table*}[htb]
\footnotesize
\center
\caption{Summary of the tested data sets for the VRPRDL.\label{tab:comparison_VRPRDL}}
\begin{threeparttable}
{\begin{tabular}{@{\extracolsep{2pt}} lrr rrr rr}\toprule
Instance set  & $|V_C|$& \cite{Ozbaygin.2017} &\multicolumn{2}{c}{\cite{Dumez.2021_trb}}& \cite{Tilk.2020} &\multicolumn{2}{c}{ALNS} \\
\cline{3-3}\cline{4-5}\cline{6-6}\cline{7-8}\rule{0pt}{2.5ex}  
& & Best 1 & Best 5& Avg. 5 & Best 1& Best 5 & Avg. 5\\
\midrule
1-5 &15   &\textbf{\num{6072.0}}& \textbf{\num{6072.0}}&6072.0  & \textbf{\num{6072.0}} & \textbf{\num{6072.0}} &\num{6072.0}  \\
6-10 &20   &\textbf{\num{6848.0}}& \textbf{\num{6848.0}}& \num{6848.0}& \textbf{\num{6848.0}}& \textbf{\num{6848.0}} &\num{6848.0}   \\
11-20 &30   &\textbf{\num{18595.0}}& \textbf{\num{18595.0}}& \num{18595.0} & \textbf{\num{18595.0}} & \textbf{\num{18595.0}} &\num{18595.0} \\
21-30 &60  &\textbf{\num{37213.0}}& \textbf{\num{37213.0}}& \num{37213.0}& \textbf{\num{37213.0}} & \textbf{\num{37213.0}} &\num{37213.0}  \\
31-40 &120  &\num{53881.0}& \textbf{\num{53738.0}}& \num{53738.4}&  \textbf{\num{53738.0}} & \textbf{\num{53738.0}} &\num{53738.8} \\
41-50-v1 &40  &\num{29842.0}& \textbf{\num{29838.0}}& \num{29838.0} & \textbf{\num{29838.0}}& \textbf{\num{29838.0}} &\num{29838.0}\\
41-50-v2 &40   &\textbf{\num{21863.0}}& \textbf{\num{21863.0}}& \num{21863.4}& \textbf{\num{21863.0}} &\textbf{\num{21863.0}} &\num{21864.9}  \\
\\
Sum &  &\num{174314.0} &\textbf{\num{174167.0}}& \num{174167.8} & \textbf{\num{174167.0}}& \textbf{174167.0} & \num{174169.7}  \\
$\overline{\text{T}}$[s] &  & 2961.0 & \multicolumn{2}{c}{17.3} & 148.5 & \multicolumn{2}{c}{16.9}\\
Processor& &  Xe-2.3G& \multicolumn{2}{c}{Xe-2.57G} &i7-3.5G& \multicolumn{2}{c}{i5-2.3G}\\
\bottomrule
\end{tabular}}
 \end{threeparttable}
\end{table*}

\begin{table*}[htb]
\footnotesize
\center
\caption{Summary of the tested data sets for the VRPHRDL.\label{tab:comparison_VRPHRDL}}
\begin{threeparttable}
{\begin{tabular}{@{\extracolsep{2pt}} lrr rrr rr}\toprule
Instance set  & $|V_C|$& \cite{Ozbaygin.2017} &\multicolumn{2}{c}{\cite{Dumez.2021_trb}}& \cite{Tilk.2020} &\multicolumn{2}{c}{ALNS} \\
\cline{3-3}\cline{4-5}\cline{6-6}\cline{7-8}\rule{0pt}{2.5ex}  
& & Best 1 & Best 5& Avg. 5& Best 1& Best 5 & Avg. 5\\
\midrule
1-5 &15   &\textbf{5450.0}& \textbf{5450.0}&  5450.0&\textbf{5450.0}& \textbf{5450.0} &5450.0 \\
6-10 &20   &\textbf{5604.0}& \textbf{5604.0}& 5604.0& \textbf{5604.0} & \textbf{5604.0} &5604.0 \\
11-20 &30   &\textbf{\num{15128.0}}& \textbf{\num{15128.0}}& \num{15128.0}& \textbf{\num{15128.0}} & \textbf{\num{15128.0}} &\num{15128.0}\\
21-30 &60  &\num{26829.0}& \textbf{\num{26800.0}}& \num{26800.0}& \textbf{\num{26800.0}}& \textbf{\num{26800.0}} &\num{26800.0}\\
31-40 &120  &\num{38610.0}& \textbf{\num{37252.0}}& \num{37310.4}& \num{37583.0} & \textbf{\num{37252.0}} &\num{37303.6}\\
41-50-v1 &40  &\num{27997.0}& \textbf{\num{27996.0}}& \num{27996.0}& \textbf{\num{27996.0}} & \textbf{\num{27996.0}} &\num{27996.0}\\
41-50-v2 &40   &\num{20977.0}& \textbf{\num{20958.0}}& \num{20958.0}& \textbf{\num{20958.0}}& \textbf{\num{20958.0}} &\num{20958.0} \\
\\
Sum &  &\num{142595.0} &\textbf{\num{139188.0}}& \num{139246.4} &\num{139519.0}& \textbf{\num{139188.0}} & \num{139239.6}\\
$\overline{\text{T}}$[s] &  & 6587.1* & \multicolumn{2}{c}{17.3} & 2924.1 & \multicolumn{2}{c}{17.0}\\
Processor& &  Xe-2.3G& \multicolumn{2}{c}{Xe-2.57G}&i7-3.5G& \multicolumn{2}{c}{i5-2.3G}\\
\bottomrule
\end{tabular}}
\begin{tablenotes}
 \item[*] \cite{Ozbaygin.2017} did not report solution times for the instances with $|V_C|=40$.
 \end{tablenotes}
 \end{threeparttable}
\end{table*}

\subsubsection{New Instances for the VRPTWTF and FSMTWTF}

We propose new instances based on modifications of the classical instances for the VRPTW from \cite{Solomon.1987} to analyze our performance with respect to the VRPTWTF and FSMTWTF. These instances are separated into three sets, that are clustered, random, and semi-clustered, and denoted as C, R, and RC. To use these classical instances for our problem with transshipment facilities, we choose a similar approach as \cite{Baldacci.2017} and derive our instances from the LRP literature. More specifically, for the transshipment facilities, we use the depots 2-6 from the LRPTW benchmark instances of \cite{Koc.2016}, who used a discrete uniform distribution to draw additional depot locations for each of the instance sets from \cite{Solomon.1987}. The central depot coordinates remain unchanged from the original Solomon instances. 

With regard to the time windows and service times, the instances are adjusted as follows:

\begin{itemize}
\item The customer location time windows are shifted forward in time by adding the service times from the Solomon instances to the start and end times of the time windows of the VRPTW instances. This way we can model the original service times as location preparation times so that the time spent at a transshipment facility is independent of the number of customer requests. Furthermore, transshipment facilities have the same location-dependent preparation times $t^p_i$ as the customer locations. 

\item Transshipment facilities can only be visited early during the routes so that the latest possible arrival time of all transshipment facilities equals the rounded maximum travel time between the depot $v_0$ and any transshipment facility ($tw_i^S =0$ and $tw_i^E = \lceil max_{i=v_0, j \in V_F}(t_{ij})\rceil$). 

\end{itemize}

Although the VRPTW typically has a hierarchical objective function of reducing the number of vehicles first and the distance second, we refrain from doing so for the VRPTWTF, as the fleet minimization objective favors the transshipment facilities too strongly, so that the routes consist almost exclusively of visiting transshipment facilities. Instead, for each instance of the VRPTWTF, we limit the number of available vehicles to the minimum number of vehicles reported for the original instances of the VRPTW in the literature. This also gives us information on the best solution values without using any of the transshipment facilities for each instance and allows us to assess how the addition of transshipment facilities can impact costs.

The instances for the FSMTWTF are also based on the Solomon instances and use the same modifications regarding the time windows and service times. For the vehicles and their corresponding costs and capacities, we use the vehicles and cost structure A from the FSMTW instances of \cite{Liu.1999}. Regarding the maximum number of iterations, we set $\eta^{max} =\num{35000}$ because it proved to be a good trade-off between run time and solution quality. 

In order to validate our Algorithm's performance on the new FSMTWTF instances, we also tested our algorithm on instances for the FSMTW that have been studied by several authors from literature. The results, summarized in Table \ref{tab:comparison_FSMTW}, show that our ALNS is very competitive for the FSMTW, finding new best solutions for four instances (see \ref{sec:details_FSMTW}).

\begin{table*}[!htp]
\footnotesize
\center
\caption{Summary of the tested data sets for the FSMTW.\label{tab:comparison_FSMTW}}
{\begin{tabular}{@{\extracolsep{2pt}}  lr c r rrr r rrr}\toprule
Instance set & \#  & $|V_C|$& \cite{Vidal.2014} &\cite{Koc.2015}& \cite{Penna.2019} &\multicolumn{2}{c}{ALNS} \\
\cline{4-4}\cline{5-5}\cline{6-6}\cline{7-8}\rule{0pt}{2.5ex}  
& & & Best 5 & Best 5& Best 10& Best 10 & Avg. 10\\
\midrule
C1 & 9 & 100& \textbf{\num{63746.78}} & \textbf{\num{63746.78}} & \textbf{\num{63746.78}}& \textbf{\num{63746.78}} & \num{63746.78} \\
R1& 12 & 100&  \textbf{\num{48375.38}}& \num{48497.56} &  \num{48385.69} &\num{48377.28}& \num{48507.54}\\
RC1 & 8 &100& \num{39129.97} & \num{39331.28} & \num{39108.26} & \textbf{\num{39099.14}} & \num{39180.60}\\
C2 & 8 & 100 &\textbf{\num{45494.00}} &\textbf{\num{45494.00}} & \textbf{\num{45494.00}} &\textbf{\num{45494.00}}& \num{45494.2}\\
R2& 11 & 100 & \num{34671.5} & \textbf{\num{34653.26}} & \num{34672.03} & \num{34676.70} & \num{34779.66}\\
RC2 & 8 &100 & \textbf{\num{33680.83}} & \num{33681.82} & \textbf{\num{33680.83}} & \num{33687.66} & \num{33792.85}\\
\\
Sum & & & \num{265098.46} & \num{265404.69} & \num{265087.59} & \textbf{\num{265081.53}} & \num{207155.09}\\
$\overline{\text{T}}$& & &305.24 &283.60 & 193.26& \multicolumn{2}{c}{252.88}\\
Processor& & & Opt-2.2G& Xe-2.6G&i7-2.93G& \multicolumn{2}{c}{i5-2.3G}\\
\bottomrule
\end{tabular}}
\end{table*}

Table~\ref{tab:comparison_vrptwtf} and ~\ref{tab:comparison_fsmtwtf} give a summary of the results obtained for the VRPTWTF and FSMTWTF, highlighting the usage of the transshipment facilities, and comparing the results with the best-known solution values of the VRPTW and FSMTW. Furthermore, Tables~\ref{tab:VRPTWTF} and \ref{tab:FSMTWTF} in  \ref{sec:details_vrptw} and \ref{sec:details_FSMTWTF} provide detailed results. The results show that for many of the instances, the number of vehicles, as well as the objective value, can be reduced, compared to the results of the VRPTW without transshipment facilities (column Avg. Gap. \%). On average, the usage of transshipment facilities leads to a cost reduction of 4.08~\% for the VRPTWTF. However, the potential for cost reductions seems to depend strongly on the instances. For example, for the clustered instances (C) of the VRPTWTF no improvements are found by adding transshipment facilities. Furthermore, the results show that fewer transshipment facilities are used for the FSMTWTF than the VRPTWTF, and smaller cost improvements compared to the FSMTW are observed. Especially for the instances with longer planning horizons (C2, R2, and RC2), only a few improvements over the best solutions without transshipment facilities are found for the FSMTWTF. The relatively small cost savings on the FSMTWTF instances can be explained by the fact that the instances have, in contrast to the VRPTW, no fleet size restrictions and as such make it easier to find good solutions without using transshipment facilities.

\begin{table*}[!htp]
\footnotesize
\center
\caption{Summary of the tested data sets for the VRPTWTF.\label{tab:comparison_vrptwtf}}
{\begin{tabular}{ lrr rrrr r}\toprule
Instance set & \# & $\overline{\text{\#r}}$ &  $\overline{\text{\#f}}$ &  $\overline{\text{\#c}}$ & $\overline{\text{z*}}$&  $\overline{\text{T}}[s]$ & Avg. Gap. \%\\
\midrule
C1 & 9 & 10.00 &0.00 &0.00 & 828.38 &84.69& ~0.00\\
R1& 12 & 10.92 & 2.83 & 15.50 & 1080.62 &173.88 & -10.14\\
RC1 & 8 & 10.88 & 3.38 & 14.88 & 1271.40 &122.03& -7.39 \\
C2 & 0 & 3.00 &0.00 &0.00 & 589.86& 132.04 & ~0.00\\
R2& 0 & 2.73 & 0.82 & 4.36 & 929.49 & 312.52 &-2.15\\
RC2 & 0 & 3.25 & 1.50 & 9.38 & 1081.76 &247.3 & -2.99\\
\bottomrule
\end{tabular}}
\end{table*}

\begin{table*}[!htp]
\footnotesize
\center
\caption{Summary of the tested data sets for the FSMTWTF.\label{tab:comparison_fsmtwtf}}
{\begin{tabular}{ lrr rrrr r}\toprule
Instance set & \# & $\overline{\text{\#r}}$ &  $\overline{\text{\#f}}$ &  $\overline{\text{\#c}}$ & $\overline{\text{z*}}$&  $\overline{\text{T}}[s]$ & Avg. Gap. \%\\
\midrule
C1 & 9 & 19.00 &1.00 &1.00 & 7079.06 &265.08 &-0.06\\
R1& 12 & 19.08 & 2.83 & 8.67 & 4013.05 &313.97 & -0.42\\
RC1 & 8 & 15.25 & 2.00 & 7.75 & 4859.92 &400.23& -0.54 \\
C2 & 8 & 5.00 &0.00 &0.00 & 5686.75& 396.01& ~0.00\\
R2& 11 & 5.00 & 0.55 & 2.09 & 3147.23 & 427.94&-1.94\\
RC2 & 8 & 11.30 & 0.13 & 1.30 & 4213.26 &430.28 & ~0.08\\
\bottomrule
\end{tabular}}
\end{table*}

\subsection{Real-World Instances}\label{sec:real-world_instances} 
In this section, we give an overview of the real-world instances newly created and the computational results obtained. First, we describe how the real-world instances were created and give an overview of the attributes and data sources. Second, we report our computational results on these instances and conduct a sensitivity analysis on the impact of the transshipment costs, demand size, time windows, and vehicle fleet on the transshipment decisions. All of the newly created instances are available upon request.

\subsubsection{Description}
\begin{figure}[htb]
{
\centering
\includegraphics[height=8.0cm,keepaspectratio]{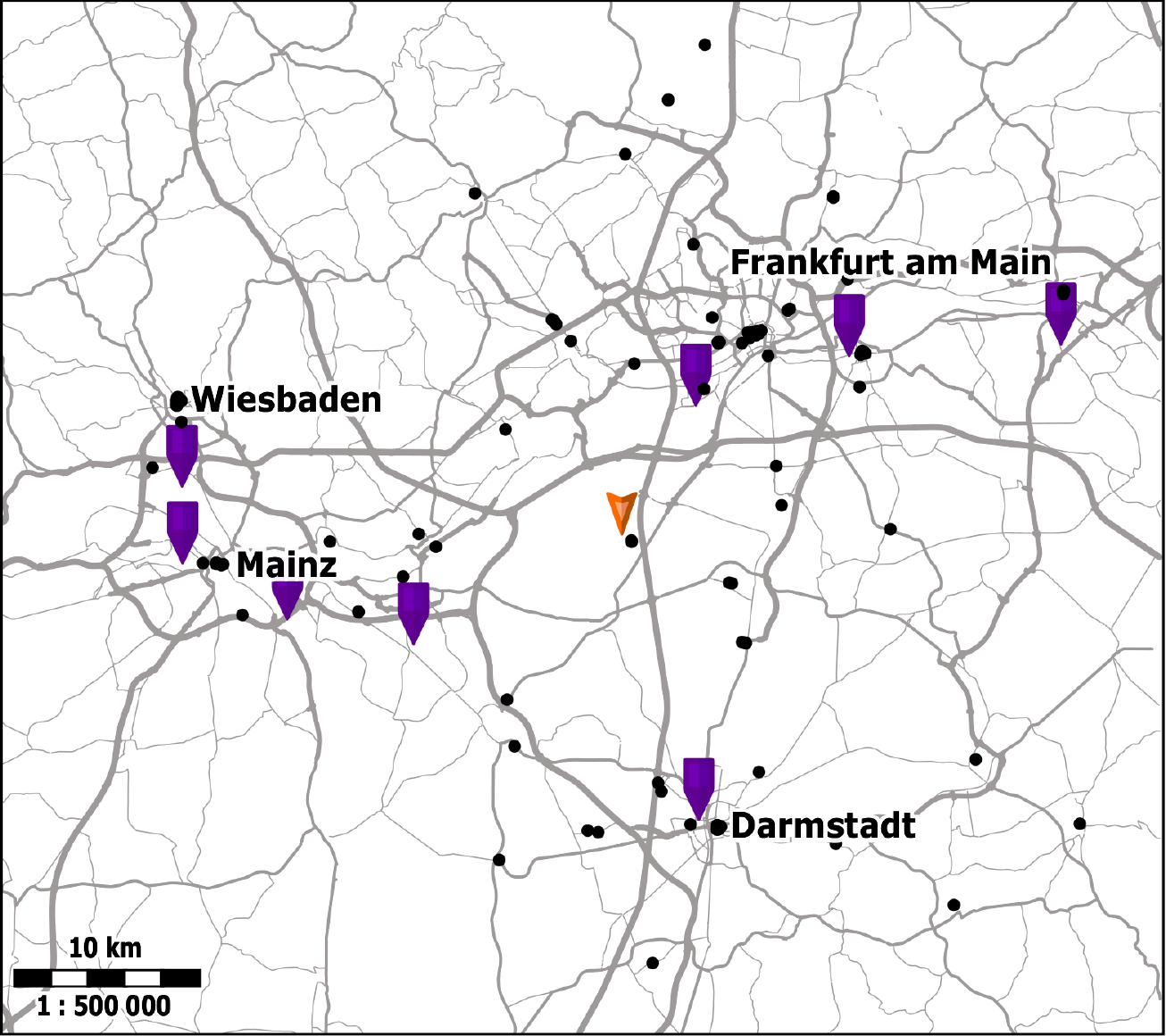}
\caption{Map of the Frankfurt Rhine-Main area with the central depot (orange arrow), transshipment facilities (purple squares), and customer locations (black dots) (Map data copyrighted OpenStreetMap Contributors).\label{fig:map-real-world}}
}
\end{figure}

The instances are based on the Frankfurt-Rhine-Main metropolitan area in Germany and feature either 50 or 100 customers, eight transshipment facilities, and one central depot. The customer locations for both instance sizes are based on the real geographic coordinates of retail stores. The central depot, as well as the transshipment facilities, are based on the geographic coordinates of real facilities from LSPs and are the same for both instances. To obtain a realistic mix of customer locations within urban and non-urban areas, we retrieved the retail locations from 13 retail chains within a 30 km radius of the central depot and selected two subsets of these locations. These retail chains operate between 4 to 20 stores within the area of which 10~\% to 100~\% are located in urban centers. For the transshipment of goods at transshipment facilities, we assume that only the customer requests of customers located in one of the seven largest cities of the Frankfurt-Rhine-Main metropolitan area can be transshipped. For this purpose, we include eight potential transshipment facilities. Each urban area has either one or two transshipment facilities and, except for one transshipment facility, all serve only one urban area. For both instances, 60~\% of the customers can be transshipped at a transshipment facility. Figure~\ref{fig:map-real-world} gives an overview of the geographic distribution of all the locations.

For the distances and travel times between locations, we retrieved the distances and travel times for each pair of locations both ways, using the HERE Routing API v7, with the mode specified as \textit{truck} and \textit{fastest} and \textit{truckType=truck}. In this context, it is worth noting that due to the presence of truck transit bans in the region, some triangle inequality violations occur. This means that visiting a sequence of customer locations $\{v_1, v_2, v_3\}$, with $v_2$ being in a transit ban zone, might be shorter than the sequence $\{v_1, v_3\}$, as vehicles with at least one delivery within the transit ban zone are exempt from the ban.

While we keep the subset of locations constant for the two instance sizes, we vary the time window lengths, customer demands, transshipment costs, and vehicle fleet during our experiments to analyze how they affect the transshipment decisions. Table~\ref{tab:summary_real-world} gives an overview of the key characteristics of the instances created for our experiments, which we will describe in the following.

\begin{table}[!htb]
\footnotesize
\centering
\caption{Real-world instance and experiment summary.\label{tab:summary_real-world}}
\begin{tabular}{lr}
\toprule
Factor & Values\\
\midrule
$|V_C|$ & $\{50,100\}$\\
$|V_F|$ &8\\
$\bar{q}$ &$[1,5]$\\
$t^p_i$ &15\\
$t^s_i$ &$\bar{q}$ \\
$\kappa$  & $\{60, 120, 180\}$\\
$c_{tf}$ &$[1,10]$\\
Fleet &  $\{$homogeneous, heterogeneous$\}$ \\ 
\bottomrule
\end{tabular}
\end{table}

The time windows of the customer locations are set according to the following rules:
\begin{itemize}
\item Customers in urban areas have to be visited in the morning due to access restrictions in the pedestrian zones so that $[tw_i^S,tw_i^E]$ such that $30+t^p_i\leq tw_i^S \leq 90+t^p_i$ and $tw_i^E=tw_i^S +\kappa$. 
\item Transshipment facilities should be visited in the morning, so that $tw_i^S =0$ and $tw_i^E = 120$.
\item For other customers that cannot be served by a transshipment facility, the time windows $[tw_i^S,tw_i^E]$ are set randomly, so that $t^p_i\leq tw_i^S \leq 450$ and $tw_i^E=tw_i^S +\kappa$. 
\end{itemize} 

Depending on the instance, we assume for the time window length $\kappa$ values of 60, 120, and 180. To analyze the impact of demand size $q_i$ on the transshipment facility usage, we systematically vary the demand for all customers in the range of 1 to 5 pallets. We modify the demand equally for each customer ($q_i = \bar{q}$), so that the impact of demand can be analyzed more easily. In this context, we assume that the service times of customer requests correspond to the demand quantity ($t^s_i=q_i$) while the demand independent preparation times $t^p$ are assumed to be 15 minutes for all locations.

Following the literature on UCCs (see e.g. \cite{Firdausiyah.2019}) and the reported real-world UCCs \citep{Janjevic.2017}, we assume that the transshipment costs are based on the quantity transshipped and thus model a price per pallet cost scheme for the transshipment facilities. Subsequently, we calculate the costs for transshipping a customer request $cr_i$ at a transshipment facility by multiplying the cost factor per unit $c_{tf}$ with the quantity $q_i$ of $cr_i$. For our sensitivity analysis, we vary the transshipment costs per unit $c_{tf}$ in the range of $[1,10]$. Regarding the vehicle fleets, we test both a homogeneous fleet and a heterogeneous fleet consisting of three vehicle types. For the homogeneous fleet, we assume a single medium-sized vehicle type with a capacity of 18 demand units. For the heterogeneous fleet, we add two additional vehicle types -- one smaller and one larger -- with capacities of 12 and 34 demand units, respectively. Each of the three vehicle types has its distance-based and fixed usage cost, while the duration-based costs are assumed to be independent of the vehicle type. Table~\ref{tab:vehicles-real-world} summarizes the three vehicle types.

\begin{table}[!htb]
\footnotesize
\centering
\caption{Vehicle properties for the real-world instance.\label{tab:vehicles-real-world}}
\begin{tabular}{lrrr}
\toprule
Vehicle property & Small Truck ($k=1$) & Medium Truck ($k=2$) & Large Truck ($k=3$)\\
\midrule
Distance-based cost $c_k^d$ [per km]& 0.33 & 0.46 &  0.61\\
Duration-based cost $c_k^t$ [per h]& 21& 21 &21\\
Fixed cost $c_k^f$ [per tour]& 55.35& 72.01 & 118.33\\
Capacity $Q_k$& 12 & 18 &34\\
\bottomrule
\end{tabular}
\end{table}

\subsubsection{Results and Managerial Insights}
\begin{figure}[!htb]
\footnotesize
\centering
\includegraphics[width=\textwidth,keepaspectratio]{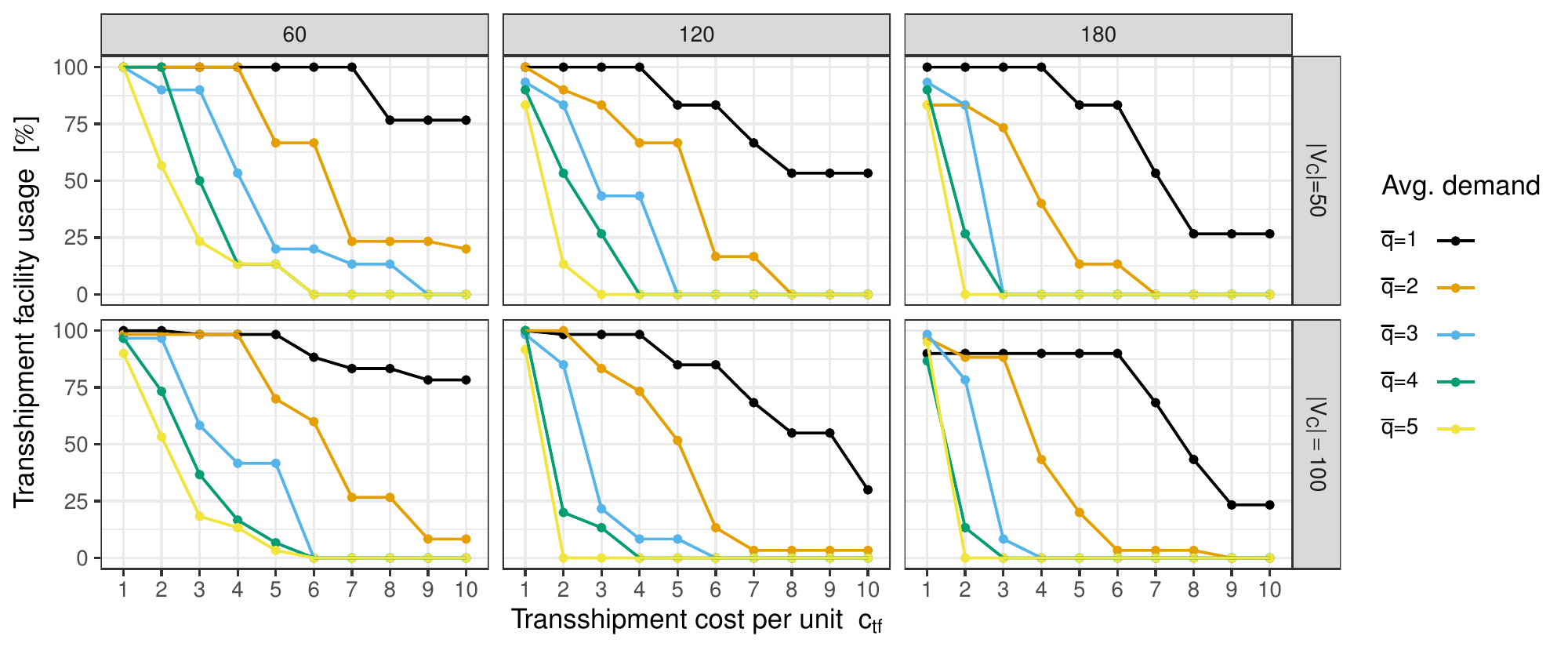}
\caption{Analysis of the impact of time window length $\kappa$ (columns), number of customers $|V_C|$, demand $\bar{q}$, and transshipment costs per unit $c_{tf}$ on the percentage of transshipment facility usage when using a heterogeneous fleet.}\label{fig:ucc_usage_het}
\end{figure}

\begin{figure}[!htb]
\footnotesize
\centering
\includegraphics[width=\textwidth,keepaspectratio]{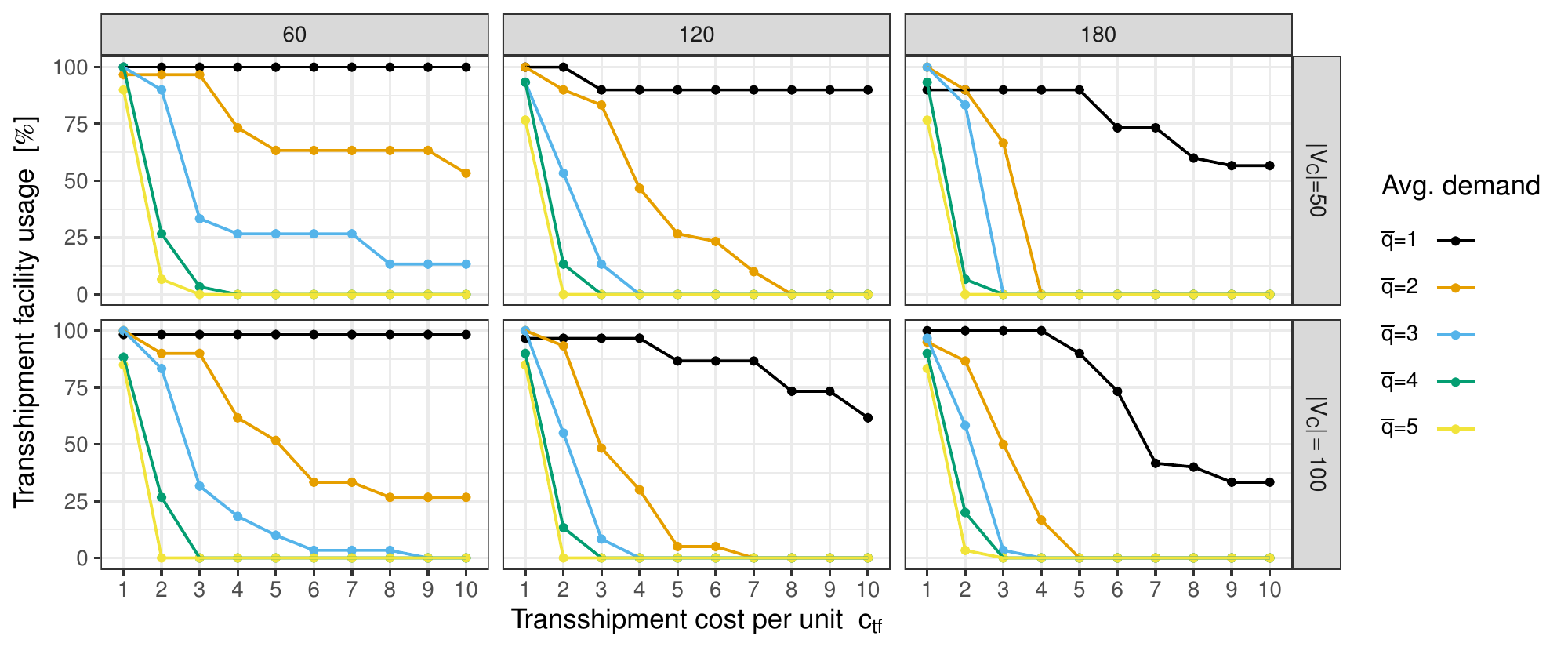}
\caption{Analysis of the impact of time window length $\kappa$ (columns), number of customers $|V_C|$, demand $\bar{q}$, and transshipment costs per unit $c_{tf}$ on the percentage of transshipment facility usage when using a homogeneous fleet consisting only of medium-sized trucks.}\label{fig:ucc_usage_hom}
\end{figure}

In this section, we present the results of our experiments to quantify the impact of transshipment costs, demand size, time windows, number of customers, and vehicle fleet on the transshipment decisions, cost, and fleet mix. Figure~\ref{fig:ucc_usage_het} provides the percentage of transshipment facility usage for the experiments with the heterogeneous fleet, and Figure~\ref{fig:ucc_usage_hom} the results for the experiments with the homogeneous fleet. AS expected, the results show that, in general, the use of the transshipment facilities declines as the usage costs per unit $c_{tf}$ increase. Moreover, the results indicate that the demand $\bar{q}$ has a strong impact on the transshipment facility use. With increasing demand per customer request, the percentage of customer requests transshipped rapidly decreases. This relationship can be explained by the quantity-based pricing of the transshipment facilities, which reduces the potential cost savings (e.g., time and distance savings) at higher transshipment quantities per stop. It is also in line with previous research from \cite{Janjevic.2017}, who showed that the cost-attractiveness for using a UCC strongly decreases as the number of pallets per stop increases.

As both figures show, the length of the time windows is another important influencing factor. With smaller time window lengths $\kappa$, the use of the transshipment facilities becomes cost-attractive and increases. In particular, for a homogeneous fleet with a narrow time window length of $\kappa=60$, the percentage of customer requests transshipped remains consistently high across the studied transshipment cost levels. Wider time windows, on the other hand, show a stronger price sensitivity so that the fraction of customer requests transshipped decreases more as transshipment costs increase.

When comparing the transshipment facility usage between the homogeneous and heterogeneous fleet, it can be observed that for customer demands of $\bar{q} =1 $ the percentage of customer requests transshipped seems to be higher for the homogeneous fleet than for the heterogeneous fleet. However, at larger demand sizes ($\bar{q}>=2$), the fraction of customer requests transshipped is overall slightly higher for the heterogeneous fleet than the homogeneous fleet. This could be explained by the fact that in the heterogeneous case, the large trucks ($k = 3$) can load more customer requests to be delivered to the transshipment facilities, making the deliveries to the transshipment facilities more efficient and allowing for larger cost savings.

The comparison of the two instance sizes does not show a clear relationship. For runs with narrow time windows, a homogeneous fleet, and customer demand $\bar{q} \leq 3$, the fraction of orders transshipped is higher for the instance with 50 customers than for the instance with 100 customers. However, this relationship between the number of customers and the fraction of orders transshipped seems to be weaker when a heterogeneous fleet is employed and the time window length is longer.

\begin{figure}[!htb]
\footnotesize
\centering
\includegraphics[width=\textwidth,keepaspectratio]{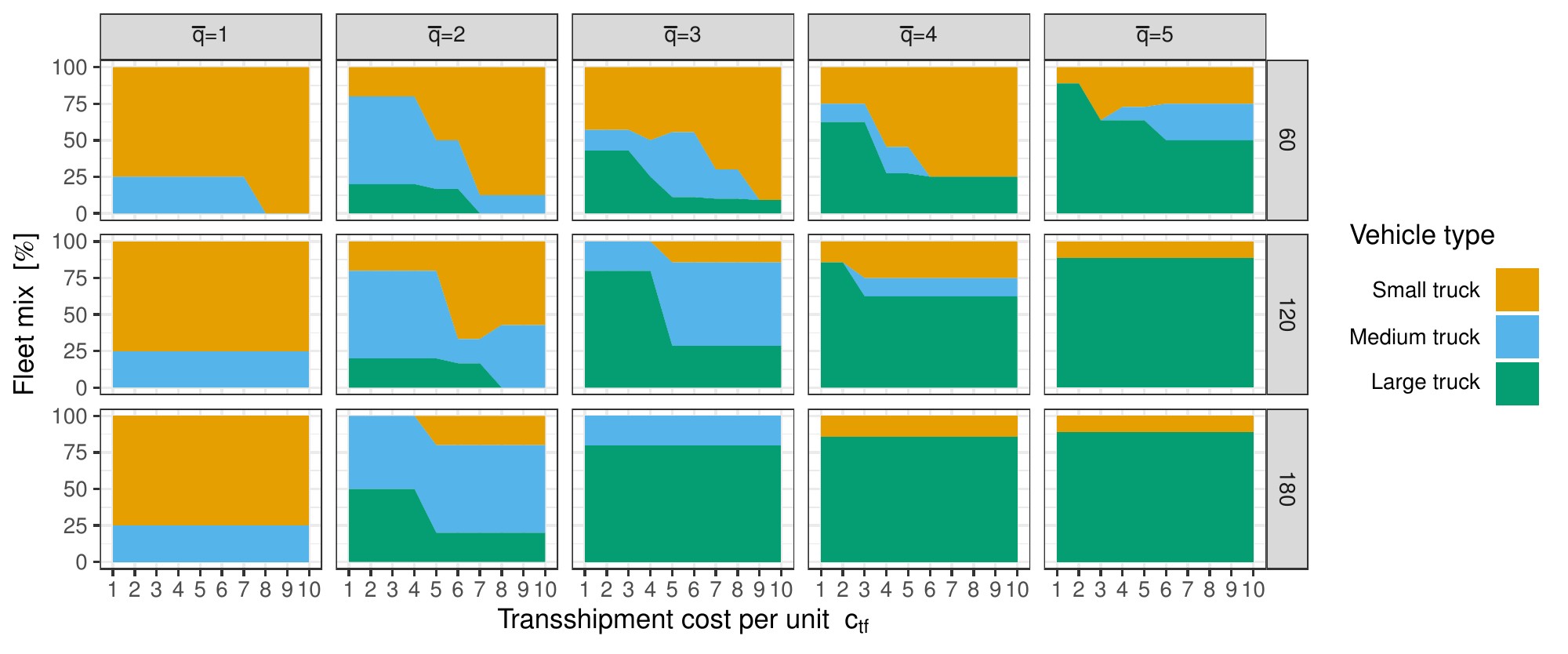}
\caption{Analysis of the impact of time window length $\kappa$ (rows), demand $q$, and transshipment costs per unit $c_{tf}$ on the fleet mix for $|V_C|=50$.}\label{fig:fleet_mix_50}
\end{figure}

The analysis of the resulting fleet mix for the instance with $|V_C|=50$, presented in Figure~\ref{fig:fleet_mix_50}, and for $|V_C|=100$ in Figure~\ref{fig:fleet_mix_100} in \ref{sec:real-world-figs}, shows how the vehicle types deployed depend on the demand per customer, time windows, and transshipment cost. First, it can be observed that the demand sizes seem to have a strong impact. With low demand per customer ($\bar{q} \in \{1,2\}$), the small and medium-sized trucks are preferred, while for higher values of $\bar{q}$ the large-sized trucks are used. Second, the figures show that with narrow time windows, smaller vehicles are selected more frequently. Third, with lower transshipment cost, and a consequently higher fraction of customer requests transshipped, the use of medium and large-sized vehicles increases for the demand sizes $\bar{q} \leq 4 $. Finally, the comparison between instances for 50 and 100 customers shows that with the larger number of customers, not only more but also larger vehicles are employed. This may be related to the fact that the higher number of customers is also accompanied by a higher customer density and thus multiple customer requests can be more easily combined into one route.

\begin{figure}[!htb]
\footnotesize
\centering
\includegraphics[width=\textwidth,keepaspectratio]{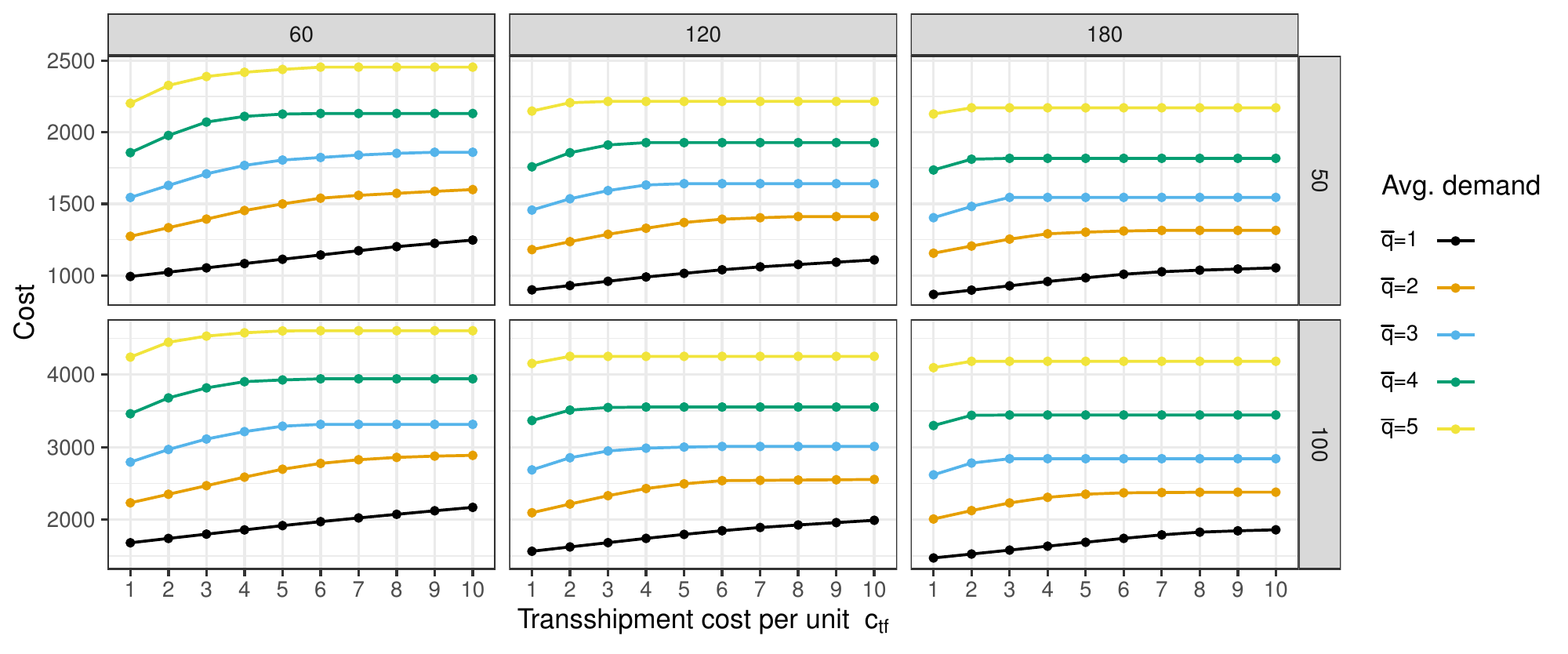}
\caption{Analysis of the impact of time window length $\kappa$ (columns), number of customers $|V_C|$, demand $q$, and transshipment costs per unit $c_{tf}$ on the costs when using a heterogeneous fleet.}\label{fig:ucc_cost_het}
\end{figure}

Lastly, we study how the overall solution costs are impacted by the factors studied. Figure~\ref{fig:ucc_cost_het}, as well as Figure~\ref{fig:ucc_cost_hom} in \ref{sec:real-world-figs}, provide an overview, of how the transshipment costs per unit $c_{tf}$ and customer demand impact the overall solution cost, separated by the two instance sizes and three time window lengths. As the transshipment costs per unit increase and thus the transshipment facility usage decreases, the overall costs per factor level combination increase, up to the point where no transshipment facility is used.

From the results of our analysis, we conclude that the decision of whether to transship customer requests at a transshipment facility depends largely on the price and demand per customer request. In addition, it becomes apparent that due to the quantity-based pricing of transshipment facilities, LSPs with low demand per customer or stop would be the primary target group for transshipment facilities. Indeed, this group of LSPs has also been previously identified by \cite{Browne.2005} as the potential beneficiaries of UCCs. As a consequence, initiatives a using quantity-based scheme should focus on attracting LSPs with low drop sizes per stop. Alternatively, different pricing models to attract LSPs with larger demands per stop should be developed.

Time windows, especially at the narrow time window width $\kappa=60$ min, can increase the potential cost-benefits of using a transshipment facility. This is also consistent with other studies from the literature, which have shown that time windows must be particularly narrow to influence the decision to use transshipment facilities, such as UCCs (\cite{Elbert.2018}, \cite{vanHeeswijk.2020}). Consequently, local governments could support the use of transshipment facilities through narrow time windows. Moreover, the comparison of the two instance sizes shows that the time window widths seem to have a stronger influence when the number of customers to be visited is lower.

Concerning the fleet mix, it can be seen that depending on the demand and usage costs, but also on the time window length, the composition of the fleet changes. Especially, in the case of medium demand per customer, the share of medium and large vehicles increases when transshipment facilities become more cost-attractive. This suggests that LSPs that want to regularly use transshipment facilities for a portion of their customer requests should ideally adjust their fleet accordingly. At the same time, however, this also means that the benefits of transshipment facilities can vary greatly depending on the LSP's fleet already in place.

\section{Conclusion and Further Research}\label{sec:conclusion}
In this article, we presented an ALNS with an embedded local search for different vehicle routing problems with transshipment facilities. The VRPTF and its variants describe an extension to the VRP where, for a fee, some customer requests can be delivered to third-party transshipment facilities, such as urban consolidation centers, instead of being delivered directly. Thereby, we extended the existing approaches on vehicle routing problems with transshipment facilities to include both time windows and heterogeneous fleets. Furthermore, we introduced new operators for removal, as well as local search moves, and insertion diversification methods that consider transshipment facilities.

To test our algorithm on various problem instances and assess the importance of different algorithmic components, we tuned the parameter of our ALNS using the irace package. Subsequently, we tested our algorithm on several instances from the literature as well as new instances derived from classical VRPTW instances and real-world data. We also used related problems, such as the VRPRDL, VRPHRDL, and FSMTW to validate the performance of our algorithm. Furthermore, we introduced new instances for the VRPTWTF and FSMTWTF, and provided solutions for these, and compared the results to the normal versions without transshipment facilities.

In comparison to the results from the literature, our algorithm seems to be very competitive. For the VRPTF our ALNS outperforms the existing exact algorithm, requiring only a fraction of the computational time. As a result, we were able to find new best solutions for 38 out of 65 VRPTF instances from the literature. For the VRPRDL and VRPHRDL, the best-known solutions reported in the literature were also found. Likewise, our algorithm also shows promising results on the well-studied FSMTW instances from the literature, obtaining new best results for four instances.

The computational results on our real-world instance, using a mixed cost function that considers distance-dependent and time-dependent costs, as well as fixed costs, show that many factors influence the decision of whether to transship customer requests at a third-party transshipment facility. In particular, the demand per customer, and the transshipment fees charged by the third-party subcontractor, but also the length of the customer time windows have a large impact on the transshipment decisions.

For future research, more factors influencing the transshipment decisions, such as the customer structure and density could be investigated using simulation. Moreover, the ALNS could be extended by considering additional real-world characteristics, such as vehicle-dependent toll schemes and travel times, incompatibility among goods, and driving and working hour regulations.

\section*{CRediT authorship contribution statement}
\textbf{Christian Friedrich:} Conceptualization, Methodology, Software, Validation, Investigation, Data curation, Formal analysis, Writing -- original draft, Writing -- review \& editing, Visualization. \textbf{Ralf Elbert:} Conceptualization, Supervision, Project Administration, Writing – Review \& Editing.

\section*{Funding}
This research did not receive any specific grant from funding agencies in the public, commercial, or not-for-profit sectors.


\appendix
\section{Detailed Results}
\subsection{VRPTF Instances}\label{sec:details_vrptf}

\begin{table*}[!h]
\footnotesize
\center
\caption{Detailed results on set (I), based on \cite{Akca.2009}.\label{tab:comparison_baldacci_akca}}
\begin{threeparttable}
{\begin{tabular*}{\textwidth}{ @{\extracolsep{\fill}} lcc rrc rll lll r}\toprule
\multirow{2}{1.8cm}{Instance}& \multirow{2}{0.8cm}{$|V_C|$} & \multirow{2}{0.8cm}{$|V_F|$}   & \multicolumn{2}{l}{\cite{Baldacci.2017}} & \multicolumn{7}{c}{ALNS}\\
\cline{4-5}\cline{6-12}\rule{0pt}{2.5ex}  
&&& $z*$ & T &$z*$ & $\overline{z}$ & $\overline{\text{T}}$ &\#r&\#f&\#c &Gap[\%]\\
\midrule
cr30x5a-1&30&5 &621.00& 8.0 &  621.00& 621.00& 3.9 & 5 &0 &0 & ~0.00* \\
cr30x5a-2&30&5 &665.00& 20.0 &  665.00& 665.00& 4.8 & 5 &1 &2 & ~0.00* \\
cr30x5a-3&30&5 &575.00& 29.0 &  575.00& 575.00& 4.5 & 5 &1 &2 & ~0.00* \\
cr30x5b-1&30&5 &727.00& 10.0 &  727.00& 727.00& 4.4 & 5 &1 &1 & ~0.00* \\
cr30x5b-2&30&5 &826.00& 79.0 &  826.00& 826.00& 4.1 & 6 &0 &0 & ~0.00* \\
cr30x5b-3&30&5 &788.00& 1061.0 &788.00& 788.00& 4.5 & 7 &1 &1 & ~0.00* \\
cr40x5a-1&40&5 &738.00& 82.0 &  738.00& 738.00& 5.1 & 7 &3 &8 & ~0.00* \\
cr40x5a-2&40&5 &786.00& 3615.0 &  786.00& 787.20& 5.7 & 6 &4 &4 & ~0.00 \\
cr40x5a-3&40&5 &807.00& 3631.0 &  807.00& 810.20& 5.1 & 6 &1 &1 & ~0.00 \\
cr40x5b-1&40&5 &964.00& 49.0 &  964.00& 964.0 & 5.5 & 8 &0 &0 & ~0.00* \\
cr40x5b-2&40&5 &901.00& 34.0 &  901.00& 901.00 & 5.4 & 8 &2 &3 & ~0.00* \\
cr40x5b-3&40&5 &887.00& 81.0 &  887.00& 887.00 & 5.3 & 8 &2 &5 & ~0.00* \\
\bottomrule
\end{tabular*}}
\begin{tablenotes}
 \item[*] Optimality proven by \cite{Baldacci.2017}.
 \end{tablenotes}
 \end{threeparttable}
\end{table*}

\begin{table*}[h!]
\footnotesize
\center
\caption{Detailed results on set (II), based on \cite{Prins.2004}.\label{tab:comparison_baldacci_prodhon}}
\begin{threeparttable}
{\begin{tabular*}{\textwidth}{ @{\extracolsep{\fill}} lrr r r r rrr lllr}\toprule
\multirow{2}{1.8cm}{Instance} & \multirow{2}{0.8cm}{$|V_C|$} & \multirow{2}{0.8cm}{$|V_F|$}  & \multicolumn{2}{l}{\cite{Baldacci.2017}} & \multicolumn{7}{c}{ALNS}\\
\cline{4-5}\cline{6-12}\rule{0pt}{2.5ex}  
& &&$z*$ & T &$z*$ & $\overline{z}$ & $\overline{\text{T}}$  &\#r&\#f&\#c &Gap[\%]\\
\midrule
ppw-20-5-2-a& 20 &5 &247.00& 14.0 &  247.00& 247.00& 1.9 & 5 &1 &3 & ~0.00* \\
ppw-20-5-2-b& 20 &5 &189.00& 5.0 &  189.00& 189.00& 2.1 & 3 &1 &2 & ~0.00* \\
ppw-50-5-2-a& 50 &5 &587.00& 126.0 &  587.00 & 587.00& 7.2 & 12 &1 &2  & ~0.00*\\
ppw-50-5-2-b& 50 &5 &357.00& 3625.0 &  357.00& 357.00& 8.6 & 6 &0 &0  & ~0.00\\
ppw-50-5-3-a& 50 &5 &586.00& 3630.0 &  583.00& 583.20& 8.3 & 12 &1 &3 & -0.51\\
ppw-50-5-3-b& 50 &5 &381.00& 3644.0 &  381.00& 381.00& 9.8 & 6&0 &0  &~0.00\\
ppw-100-5-2-a& 100 &5 &1010.00& 3735.0 &  996.00& 997.80& 39.8 & 23 &1 &3 & -1.41 \\
ppw-100-5-2-b& 100 &5 &569.00& 3783.0 &  558.00& 560.30& 41.7 &11 &0&0  &-1.97\\
ppw-100-5-3-a& 100 &5 &1068.00& 3686.0 &  1065.00& 1066.80 & 36.6 & 23 &2 &3 & -0.28\\
ppw-100-5-3-b& 100 &5 &612.00& 3733.0 &  607.00& 608.30&38.3 & 11 &0 &0 & -0.82\\
ppw-100-10-2-a& 100 &10 &1030.00& 3853.0 &  1012.00& 1018.30& 41.6 & 24 &2 &5 &-1.78\\
ppw-100-10-2-b& 100 &10 &582.00& 3809.0 &  574.00& 577.10& 70.7 & 11 &0 &0 & -1.39\\
ppw-100-10-3-a& 100 &10 &1055.00& 424.0 &  1038.00& 1039.50& 40.6 & 24&3 &5 & -1.54$\dagger$\\
ppw-100-10-3-b& 100 &10 &608.00& 3773.0 &  599.00& 600.20& 80.1 & 11 &2 &3 & -1.50\\
\bottomrule
\end{tabular*}}
\begin{tablenotes}
 \item[*] Optimality proven by \cite{Baldacci.2017}.
 \item[$\dagger$] \cite{Baldacci.2017} report their value of 1055 to be optimal. We assume this to be an error.
 \end{tablenotes}
 \end{threeparttable}
\end{table*}

\begin{table*}[t!]
\footnotesize
\center
\caption{Detailed results on instance set (III), based on different authors.\label{tab:comparison_baldacci_others}}
\begin{threeparttable}
{\begin{tabular*}{\textwidth}{@{\extracolsep{\fill}} lrr r r r r rll lll}\toprule
\multirow{2}{1cm}{Instance} & \multirow{2}{0.8cm}{$|V_C|$} & \multirow{2}{0.8cm}{$|V_F|$}   & \multicolumn{2}{l}{\cite{Baldacci.2017}} & \multicolumn{7}{c}{ALNS}\\
\cline{4-5}\cline{6-12}\rule{0pt}{2.5ex}  
& &&$z*$ & T &$z*$ & $\overline{z}$ & $\overline{\text{T}}$  &\#r&\#f&\#c  &Gap[\%]\\
\midrule 
Christ-50x5& 50& 5& 514.00& 130.0 & 514.00& 519.90& 7.0 & 5 &2 &2 & ~0.00*\\
Christ-50x5B &50& 5& 533.00& 3698.0 & 530.00& 531.00& 7.2 & 5 &2 &3 & -0.57\\
Christ-75x10 &75& 10& 783.00& 3676.0 &  771.00& 776.80& 19.3 & 9 &3 &3 & -1.56\\
Christ-75x10B &75& 10&814.00& 3750.0 &  812.00& 813.00& 18.3 & 9 &4 &6& -0.25 \\
Christ-100x10 &100& 10&831.00& 3987.0 &  821.00& 823.20& 100.0 & 8 &0 &0 & -1.22 \\
Gaskell-21x5 &21& 5& 371.00& 5.0 &  371.00& 371.00& 2.1 & 4 &1 &2 &  ~0.00* \\
Gaskell-22x5 &21& 5& 554.00& 145.0 &  554.00& 550.40& 2.1 & 3 &3 &4 &  ~0.00* \\
Gaskell-29x5 &29& 5& 503.00& 683.0 &  503.00& 503.00& 4.9 & 4 &1 &1 &  ~0.00* \\
Gaskell-32x5 &32& 5 &479.00& 280.0 &  479.00& 479.00& 5.3 & 4 &1 &1 &  ~0.00* \\
Gaskell-32x5-2 &32& 5&427.00&567.0 &  427.00& 427.70& 4.9 & 3 &0 &0 &  ~0.00* \\
Gaskell-36x5 &36& 5&411.00&17.0 &  411.00& 411.40& 5.6 & 4 &1 &1 &  ~0.00* \\
Min-27x5 &27& 5&3083.00 & 24.0 &  3083.00& 3083.00& 4.5 & 4 &1 &1 &  ~0.00* \\
Perl83-12x2 &12& 2&100.00& 2.0 &  100.00& 100.00& 1.0 &8 &0 &0 &  ~0.00*\\
Perl83-55x15 &55& 15&453.00& 278.0 &  453.00& 453.40& 12.3 & 10 &3 &3 & ~0.00*\\
Perl83-85x7 &85& 7&618.00& 3736.0 &  617.00& 617.20& 20.5 & 11 &0 &0&-0.16  \\
P111112-100x10 &100& 10 &1346.00& 3846.0 &  1315.00& 1337.80& 66.1 & 11 &0 &0 &-2.36 \\
P111122-100x20 &100& 20 &1252.00& 4138.0 &  1242.00& 1255.40& 64.0 & 11 & 1 &1 &-0.81  \\
P111212-100x10 &100& 10 &1266.00& 3718.0 &  1256.00& 1259.80& 41.0 & 10 &0 &0 &-0.80  \\
P111222-100x20 &100& 20 &1338.00& 4053.0 &  1310.00& 1316.60& 60.0 & 11 &3 &3 &-2.14  \\
P112112-100x10 &100& 10 &1236.00& 3889.0 &  1229.00& 1230.8& 75.5 & 11 &3 &3 & -0.57  \\
P112122-100x20 &100& 20 &1047.00& 4177.0 &  1018.00& 1025.40& 87.4 & 10 &3 &3 &-2.85  \\
P112212-100x10 &100& 10 &892.00& 3918.0 &  877.00& 877.00 &69.10 & 11 &1 &1 &-1.71  \\
P112222-100x20 &100& 20 &1006.00& 3747.0 &  1000.00& 1002.00& 77.2 & 10 &2 &2 &-0.60  \\
P113112-100x10 &100& 10 &1158.00&4007.0 &  1149.00& 1157.60& 68.1 & 10 &0 &0 &-0.78  \\
P113122-100x20 &100& 20 &1190.00& 3914.0 &  1175.00& 1184.40& 75.8 & 11 &4 &7 &-1.28  \\
P113212-100x10 &100& 10 &1154.00& 3717.0 &  1149.00& 1150.00& 64.0 & 10 &1 &1 &-0.44  \\
P113222-100x20 &100& 20 &1078.00& 3748.0 &  1071.00& 1071.00& 79.0 & 11 &0 &0 &-0.65 \\
P131112-150x10 &100& 10 &1833.00 & 3946.0 &  1795.00& 1807.80& 53.6 & 16 &0 &0 &-2.12 \\
P131122-150x20 &150& 20 &1769.00& 4411.0 &  1756.00& 1756.00& 79.6 & 16 &2 &2 &-0.74\\
P131212-150x10 &150& 10 &1802.00& 4147.0 &  1786.00& 1790.20& 51.7& 16 &2 &3 &-0.90\\
P131222-150x20 &150& 20&1802.00& 4371.0 &  1775.00& 1783.40& 78.7 & 16 &2 &3 &-1.52 \\
P132112-150x10 &150& 10&1783.00&4815.0 &  1759.00& 1780.20& 97.1 & 16 &2 &2 &-1.36 \\
P132122-150x20 &150& 20&1541.00& 4508.0 &  1519.00& 1526.60& 114.1 & 16 &1 &1 &-1.45  \\
P132212-150x10 &150& 10&1251.00& 4073.0 &  1242.00& 1243.40& 81.3 & 16 &0 &0 &-0.72  \\
P132222-150x20 &150& 20&1184.00& 4368.0 &  1171.00& 1174.00& 97.2 & 16 &0 &0 &-1.11 \\
P133112-150x10 &150& 10&1899.00& 4434.0 &  1892.00& 1900.60& 83.6 & 16 & 1 &1 &-0.37 \\
P133122-150x20 &150& 20&1498.00& 4908.0 &  1489.00& 1499.40& 111.2 & 16 &1 &1 &-0.60  \\
P133212-150x10  &150& 10&1245.00& 4510.0 &  1236.00& 1236.80& 84.8 & 16 &0 &0 &-0.73\\
P133222-150x20  &150& 20&1551.00& 4688.0 &  1509.00& 1529.20& 104.5 & 15 &0 &0 &-2.78 \\
\bottomrule
\end{tabular*}}
\begin{tablenotes}
 \item[*] Optimality proven by \cite{Baldacci.2017}.
 \end{tablenotes}
 \end{threeparttable}
\end{table*}

 \clearpage
\subsection{VRPTWTF Instances}\label{sec:details_vrptw}

\begingroup
\captionsetup{width=\textwidth}
\renewcommand\arraystretch{0.8}
\begin{table*}[h!]
\footnotesize
\center
\caption{Detailed results on the VRPTWTF instance set based on \cite{Solomon.1987}.\label{tab:VRPTWTF}}
\begin{threeparttable}
{\begin{tabular*}{\textwidth}{@{\extracolsep{\fill}}llr rrr llr rr}
\toprule
\multirow{2}{1.8cm}{Instance}  & \multirow{2}{0.8cm}{$|V_C|$} & \multirow{2}{0.8cm}{$|V_F|$} & \multicolumn{6}{c}{ALNS} & VRPTW-&VRPTW-\\
\cline{4-9}\rule{0pt}{2.5ex}  
  && & $z*$ & $\overline{z}$ & $\overline{\text{T}}$  &\#r&\#f&\#c & BKS*& Gap [\%]  \\
\midrule
C101 &100& 5   &828.94& 828.94& 86.4 & 10 &0 &0  &828.94&  ~~0.00\tabularnewline
C102 &100& 5   &828.94& 828.94& 89.8& 10 &0 &0  &828.94&  ~~0.00 \tabularnewline
C103 &100& 5   &828.06& 828.06& 90.1 & 10 &0 &0  & 828.06& ~~0.00 \\
C104 &100& 5   &824.78& 824.78& 84.2& 10 &0 &0  &824.78& ~~0.00 \\
C105 &100& 5   &828.94& 828.94& 81.3 &10 &0 &0  &828.94&  ~~0.00 \\
C106 &100& 5   &828.94& 828.94& 78.4 &10 &0 &0  &828.94&  ~~0.00 \\
C107 &100& 5  &828.94& 828.94& 81.4 &10 &0 &0 &828.94& ~~0.00  \\
C108 &100& 5   &828.94& 828.94& 82.3 & 10 &0 &0 &828.94& ~~0.00  \\
C109 &100& 5   &828.94& 828.94& 88.3 & 10 &0 &0 &828.94& ~~0.00  \\
\midrule 
R101 &100& 5   &1400.90& 1401.65& 174.3 & 15 &4 &26 &1650.8 &  -15.14 \\
R102 &100& 5   &1242.24& 1243.05& 183.1 & 13 &3 &21 & 1486.12& -16.41 \\
R103 &100& 5   &1094.32& 1095.87& 176.2 & 12 &3 &17  &1292.68& -15.34 \\
R104 &100& 5   &958.28& 961.43& 171.8 & 9 &2 &9  &1007.31& -4.87 \\
R105 &100& 5   &1215.11& 1217.03& 170.2 & 12 &4 &24  & 1377.11& -11.76 \\
R106 &100& 5   &1130.07& 1133.63& 177.7 & 11 &2 &17 &1252.03& -9.74 \\
R107 &100& 5   &1008.94& 1011.08& 169.5 & 10 &2 &11 &1104.66& ~-8.67\\
R108 &100& 5   &920.21& 922.69& 164.2 & 9 &2 &9  &960.88&  -4.23 \\
R109 &100& 5   &1071.95& 1073.19& 180.9 & 11 &3 &16 &1194.73& -10.28 \\
R110 &100& 5   &1005.22& 1006.60& 174.8 & 10 &4 &14 &1118.84&  -10.16\\
R111 &100& 5   &996.55& 1001.73& 173.7 & 10 &2 &11 &1096.72& -9.13\\
R112 &100& 5   &923.64& 927.09& 170.1 & 9 &3 &11 &982.14& -5.96  \\
\midrule 
RC101 &100& 5   &1502.96& 1508.34& 119.9 & 12 &4 &21 &1696.95& -11.43  \\
RC102 &100& 5   &1346.32& 1352.55& 122.5 &  11 &4 &22 &1554.75& -13.41  \\
RC103 &100& 5   &1221.27& 1227.73 & 118.0 & 10 &2 &9 &1261.67& -3.20 \\
RC104 &100& 5   &1121.73& 1124.88& 116.4 & 10 &3 &6  &1135.48& -1.21\\ 
RC105 &100& 5   &1368.01& 1378.28& 123.9& 12 &4 &22 &1629.44& -16.04 \\
RC106 &100& 5   &1303.20& 1311.40& 127.4 & 11 &4 &18  &1424.73& -8.53 \\
RC107 &100& 5   &1195.13& 1199.27& 126.3 & 11 &3 &11 &1230.48&-2.87  \\
RC108 &100& 5   &1112.55& 1118.33& 121.9 & 10 &3 &10  &1139.82& -2.39 \\
\midrule
C201 &100& 5   &591.56& 591.56& 122.4 & 3 &0 &0 & 591.56& ~~0.00  \\
C202 &100& 5  &591.56& 591.56& 133.5 & 3 &0 &0 & 591.56& ~~0.00  \\
C203 &100& 5   &591.17& 591.17& 134.6 & 3 &0 &0 & 591.17& ~~0.00  \\
C204 &100& 5   &590.60& 590.60& 128.4& 3 &0 &0 & 590.60& ~~0.00  \\
C205 &100& 5   &588.88& 588.88& 145.3 & 3 &0 &0  & 588.88& ~~0.00 \\
C206 &100& 5   &588.49& 588.49& 130.9 & 3 &0 &0  & 588.49&  ~~0.00 \\
C207 &100& 5   &588.29& 588.29& 132.5 & 3 &0 &0  &588.29&  ~~0.00 \\
C208 &100& 5   &588.32& 588.32& 128.7 & 3 &0 &0  &588.32& ~~0.00\\
\midrule 
R201 &100& 10   &1197.53& 1198.25& 303.1  & 4 &2 &10 &1252.37& ~-4.38\\
R202 &100& 10   &1139.98& 1147.61& 318.1& 3 &1&9 &1191.70& ~~0.00\\
R203 &100& 10   &900.56& 904.61& 310.8 &3&1 &6  &939.50& ~-4.14 \\
R204 &100& 10   &779.21& 788.99&321.0 & 2 &1 &5 &825.52& ~-5.61  \\
R205 &100& 10   &994.43& 998.21& 294.6 & 3 &0 &0 &994.43& ~~0.00  \\
R206 &100& 10   &900.59& 903.79& 300.6 & 3 &1 &3 &906.14& ~-0.61  \\
R207 &100& 10   &874.28& 876.23& 318.9& 2 &1 &5  &890.61& ~-1.83 \\
R208 &100& 10   &725.42& 729.94& 309.8 & 2 &1 &4 &726.82& ~-0.19  \\
R209 &100& 10   &909.16& 915.05& 333.0 & 3 &0 &0 &909.16& ~~0.00  \\
R210 &100& 10   &929.70& 932.77& 314.8 & 3 &1 &3  &939.37& ~-1.03 \\
R211 &100& 10   &885.71& 886.33 & 313.0 & 2 &0 &0 &885.71& ~~0.00 \\
\midrule
RC201 &100& 10   &1333.35& 1338.71& 234.3 & 4  &3 &22 & 1406.94& ~-5.23 \\
RC202 &100& 10   &1284.22& 1294.07& 242.4 & 3 &2 &15 & 1365.65& ~-5.96  \\
RC203 &100& 10   &1033.27& 1041.34& 238.5 & 3 &1 &7 & 1049.62& ~-1.56  \\
RC204 &100& 10   &798.46& 801.95 & 240.6 & 3 &0 &0 &798.46& ~~0.00  \\
RC205 &100& 10   &1232.56& 1244.68& 227.3 & 4 &2 &10  &1297.65& ~-5.02 \\
RC206 &100& 10   &1129.83& 1142.32& 249.5 &3 &1 &10 &1146.32& ~-1.44  \\
RC207 &100& 10   &1024.08&1028.93& 256.2 & 3 &2 &10 &1061.14& ~-3.49  \\
RC208 &100& 10   &818.27& 828.63& 289.6 & 3 &1 &4  & 828.14&~-1.19 \\
\bottomrule
\end{tabular*}}
\begin{tablenotes}
 \item[*] Values taken from \cite{Nagata.2010}.
 \end{tablenotes}
 \end{threeparttable}
\end{table*}
\endgroup

 \clearpage
\subsection{FSMTW Instances}\label{sec:details_FSMTW}

\begingroup
\vfill
\begin{sideways}
\begin{threeparttable}
\footnotesize
\caption{Detailed results on the FSMTW instances (fleet A) I.\label{tab:FSMTW_I}}
\begin{tabular}{@{\extracolsep{2pt}}l lll lll lll lll lll @{}}\toprule
\multirow{2}{0.5cm}{Instance} & \multirow{2}{0.1cm}{$|V_C|$}& \multirow{2}{0.8cm}{BKS}& 
\multicolumn{3}{c}{\cite{Vidal.2014}}& \multicolumn{3}{c}{\cite{Koc.2015}}& \multicolumn{3}{c}{\cite{Penna.2019}}&\multicolumn{3}{c}{ALNS}\\
\cline{4-6} \cline{7-9} \cline{10-12} \cline{13-16}
\rule{0pt}{2.5ex}  
  &  & &$z*$ & $\overline{z}$ & $\overline{\text{T}}$&$z*$ & $\overline{z}$ & $\overline{\text{T}}$&$z*$ & $\overline{z}$ & $\overline{\text{T}}$&$z*$ & $\overline{z}$ & $\overline{\text{T}}$& Gap[\%]\\
\midrule
R101 & 100 &4314.36& 4314.36& 4322.04 & 276.60 & 4317.52 & --& 248.40 & 4314.36& 4325.76&120.56&4314.36&4316.10& 167.9 &~0.00\\
R102 & 100 &4166.28& 4166.28& 4175.05 & 361.80 &4173.84 & --& 358.80 &4166.28 &4181.25 & 124.45 & 4166.28 & 4186.78 & 111.3& ~0.00\\
R103 & 100 &4024.14& 4027.36& 4034.88 & 321.00 &4031.40 & --& 312.60 &4024.14 &4038.85 & 134.79 & 4026.23 & 4036.97 & 179.4& ~0.05\\
R104 & 100 &3936.40& 3936.40& 3938.92 & 288.60 &3946.44 & --& 247.20 &3936.55 &3948.00 & 113.83 & 3936.51& 3940.41 & 225.5& ~0.01\\
R105 & 100 &4122.50& 4122.50& 4130.71 &389.40 &4134.06 & --& 360.60 &4122.50 &4133.06 & 131.61 & 4122.50 & 4127.69& 210.5& ~0.00\\
R106 & 100 &4048.59& 4048.59& 4058.95 & 334.20 &4060.05 & --& 307.20 &4050.17 &4059.07 & 127.47 & 4047.34 & 4071.37 & 215.1& -0.03\\
R107 & 100 &3970.51& 3970.51& 3979.18 & 333.60 &3985.12 & --& 286.80 &3976.40 &3988.73 & 126.16 & 3973.68 & 3983.72 &255.4& ~0.08\\
R108 & 100 &3928.12& 3928.12& 3932.46 & 280.80 &3932.60& --& 392.40 &3928.12 &3935.45 & 108.28 & 3930.17 & 3945.00 & 250.4& ~0.05\\
R109 & 100 &4015.71& 4015.71& 4020.93 &288.00 &4024.83 & --& 367.20 &4015.71 &4023.34 & 128.21 & 4011.97 &4021.19 & 205.9& -0.09\\
R110 & 100 &3961.68& 3961.68& 3966.47 & 288.00 &3973.51 & --& 367.20 &3961.68 &4023.34 &114.65 & 3960.55 & 3967.05 & 236.7& -0.03\\
R111 & 100 &3964.99& 3964.99& 3973.49& 316.80 &3988.00& --& 307.20 &3971.90 &3989.84& 127.58 & 3968.81 & 3985.81 & 155.9& ~0.10\\
R112 & 100 &3917.88& 3918.88& 3926.32& 295.20 &3930.19 & --& 282.60 &3917.88 &3927.20 & 105.96 & 3918.88 & 3929.29 & 182.9& ~0.03\\
\midrule 
C101 & 100 &7093.45& 7093.45& 7093.45& 177.60 &7093.45 & --& 148.20 &7093.45 &7093.59 & 167.56& 7093.45 & 7093.45 & 124.1& ~0.00\\
C102 & 100 &7080.17& 7080.17& 7080.17& 128.40 &7080.17 & --& 159.00 &7080.17 &7080.17 & 134.68& 7080.17 & 7080.17 & 136.3& ~0.00\\
C103 & 100 &7079.21& 7079.21& 7079.21& 125.40 &7079.21 & --& 120.60 &7079.21 &7079.21 &119.00& 7079.21 & 7079.21 & 164.2& ~0.00\\
C104 & 100 &7075.06& 7075.06& 7075.06& 131.40 &7075.06 & --& 118.20 &7075.06 &7075.06 & 101.36& 7075.06 & 7075.06 & 165.1& ~0.00\\
C105 & 100 &7093.45& 7093.45& 7093.45& 199.80 &7093.45 & --& 159.00 &7093.45 &7093.60 & 154.47& 7093.45 & 7093.45 & 111.0& ~0.00\\
C106 & 100 &7083.87& 7083.87& 7083.87&136.80 &7083.87 & --& 130.20 &7083.87 &7083.87 &140.25& 7083.87 & 7083.87 & 151.6& ~0.00\\
C107 & 100 &7084.61& 7084.61&7084.61& 133.80 &7084.61 & --& 143.40 &7084.61 &7084.61 &142.94&7084.61& 7084.61 & 154.1& ~0.00\\
C108 & 100 &7079.66& 7079.66& 7079.66& 131.40 &7079.66 & --& 118.20 &7079.66 &7079.66 & 124.35& 7079.66 & 7079.66 & 179.9& ~0.00\\
C109 & 100 &7077.30& 7077.30& 7077.30& 122.40 &7077.30 & --& 131.40 &7077.30 &7077.30 & 107.98& 7077.30 & 7077.30 & 170.5& ~0.00\\
\midrule 
RC101 & 100 &5150.86& 5150.86& 5154.95& 312.60 &5173.47 & --& 308.40 &5150.86&5160.03 & 121.21& 5150.86 & 5151.78 & 225.3 & ~0.00\\
RC102 & 100 &4974.82& 4987.24& 5000.28& 288.60 &5018.83 & --& 255.60 &4974.82 &4999.64 & 122.10& 4974.82 & 4988.66 & 215.7& ~0.00\\
RC103 & 100 &4804.61& 4804.61& 4821.61& 424.80 &4850.20 & --& 388.20 &4804.61 &4837.40 & 122.75& 4804.61 & 4821.56 &155.8& ~0.00\\
RC104 & 100 &4717.63& 4717.63& 4724.10& 318.00 &4725.40 & --& 317.40 &4721.44 &4734.77 & 94.74& 4717.63 & 4725.97 & 223.2& ~0.00\\
RC105 & 100 &5035.35& 5035.35& 5035.76& 334.20 &5048.86 & --& 286.80 &5036.50 &5047.72 & 119.73& 5035.35 & 5041.05 & 170.5& ~0.00\\
RC106 & 100 &4921.13& 4936.74& 4944.74& 337.80 &4964.13 & --& 317.40 &4921.13 &4941.27 & 118.48& 4919.43 & 4932.82 & 173.9& -0.04\\
RC107 & 100 &4787.59& 4788.69& 4795.35& 304.80&4825.60 & --& 250.20 &4787.59 &4807.65 & 107.49& 4787.59 & 4797.27 & 165.0& ~0.00\\
RC108 & 100 &4708.85& 4708.85& 4709.09& 286.80 &4724.79& --& 277.80 &4711.31 &4726.02 &86.50& 4708.85 & 4721.49 & 158.3& ~0.00\\
\bottomrule
\end{tabular}
  \end{threeparttable}
\end{sideways}
\vfill
\restoregeometry
\endgroup
 \clearpage
\begingroup
\vfill
\begin{sideways}
\begin{threeparttable}
\footnotesize
\caption{Detailed results on the FSMTW instances (fleet A) II.\label{tab:FSMTW_II}}
\begin{tabular}{@{\extracolsep{2pt}}l lll lll lll lll lll @{}}\toprule
\multirow{2}{0.5cm}{Instance} & \multirow{2}{0.1cm}{$|V_C|$}& \multirow{2}{0.8cm}{BKS}& 
\multicolumn{3}{c}{\cite{Vidal.2014}}& \multicolumn{3}{c}{\cite{Koc.2015}}& \multicolumn{3}{c}{\cite{Penna.2019}}&\multicolumn{3}{c}{ALNS}\\
\cline{4-6} \cline{7-9} \cline{10-12} \cline{13-16}
\rule{0pt}{2.5ex}  
  &  & &$z*$ & $\overline{z}$ & $\overline{\text{T}}$&$z*$ & $\overline{z}$ & $\overline{\text{T}}$&$z*$ & $\overline{z}$ & $\overline{\text{T}}$&$z*$ & $\overline{z}$ & $\overline{\text{T}}$& Gap[\%]\\
\midrule
R201 & 100 &3446.78& 3446.78& 3446.78& 390.60 &3446.78 & --& 367.80 &3446.78 &3452.08 &320.01& 3446.78 & 3455.82 & 338.5& ~0.00\\
R202 & 100 &3297.42& 3308.16& 3308.16& 460.80 &3297.42 & --& 447.60 &3308.16 &3313.28 & 342.72& 3308.86 & 3315.73 & 275.8& ~0.35\\
R203 & 100 &3141.09& 3141.09& 3141.09& 339.00 &3141.09 & --& 368.40 &3141.09 &3143.21 & 321.32& 3141.09 &3141.19 & 310.0& ~0.00\\
R204 & 100 &3018.14& 3018.14& 3018.83& 417.60 &3018.14 & --& 376.80 &3018.14 &3019.95 &296.69& 3018.14 & 3020.78 & 367.1& ~0.00\\
R205 & 100 &3218.97& 3218.97& 3220.56& 384.00 &3218.97 & --& 382.80 &3218.97 &3228.75 & 305.08& 3218.97 & 3225.15 & 269.1& ~0.00\\
R206 & 100 &3146.34& 3146.34& 3150.61& 618.00 &3146.34 & --& 488.40 &3147.41 &3155.39 & 325.90& 3146.84 &3152.32 & 314.6& ~0.02\\
R207 & 100 &3077.36& 3077.58& 3080.64& 522.00 &3077.36 & --& 388.20 &3077.58 &3082.82 & 313.42& 3080.13 & 3111.44 & 302.9& ~0.09\\
R208 & 100 &2997.24& 2997.24& 2999.35& 322.20 &2997.25 & --& 380.40 &2997.24 &2999.97 & 282.11& 2998.70 & 3030.32 & 258.0& ~0.05\\
R209 & 100 &3119.56& 3122.42& 3123.30& 382.20 &3119.56 & --& 299.40 &3122.42 &3129.93 & 301.95& 3122.42 & 3122.42 & 269.6& ~0.09\\
R210 & 100 &3170.41& 3174.85& 3178.57& 415.80 &3170.41 & --& 328.20 &3174.31 &3181.35 & 331.93& 3174.85 & 3179.74 & 250.1& ~0.14\\
R211 & 100 &3019.93& 3019.93& 3021.67& 546.00 &3019.93 & --& 475.80 &3019.93 &3022.78 & 285.35& 3019.93 & 3024.76 &327.5& ~0.00\\
\midrule 
C201 & 100 &5695.02& 5695.02& 5695.02& 222.60 &5695.02 & --& 207.60 &5695.02 &5695.02 &345.40& 5695.02 & 5695.02 & 415.9& ~0.00\\
C202 & 100 &5685.24& 5685.24& 5685.24& 226.80 &5685.24 & --& 190.20 &5685.24 &5685.24 & 287.90& 5685.24 & 5685.24 &354.2& ~0.00\\
C203 & 100 &5681.55& 5681.55& 5681.55& 252.60 &5681.55 & --& 257.40 &5681.55 &5681.79 & 271.20& 5681.55 & 5681.55 & 345.1& ~0.00\\
C204 & 100 &5677.66& 5677.66& 5677.66& 256.20 &5677.66 & --& 238.20 &5677.66 &5677.88 &281.35& 5677.66 & 5677.66 & 346.3& ~0.00\\
C205 & 100 &5691.36& 5691.36& 5691.36& 238.80 &5691.36 & --& 207.60 &5691.36 &5691.36 & 300.58& 5691.36 & 5691.56 &357.8& ~0.00\\
C206 & 100 &5689.32& 5689.32& 5689.32& 229.20 &5689.32 & --& 178.20 &5689.32 &5689.32 & 270.71& 5689.32 & 5689.50 & 388.6& ~0.00\\
C207 & 100 &5687.35& 5687.35& 5687.35& 254.40 &5687.35 & --& 246.00 &5687.35 &5687.35 &294.10& 5687.35 & 5687.40 & 395.6& ~0.00\\
C208 & 100 &5686.50& 5686.50& 5686.50& 231.60 &5686.50 & --& 213.60 &5686.50 &5686.50 & 265.14& 5686.50 & 5686.50 & 387.1& ~0.00\\
\midrule 
RC201 & 100 &4374.09& 4374.09& 4378.21& 355.20 &4376.82 & --& 308.40 &4374.09 &4380.07 & 193.35& 4374.09 & 4402.00 & 329.6& ~0.00\\
RC202 & 100 &4244.63& 4244.63& 4244.65& 277.80 &4244.63 & --& 255.60 &4244.63 &4246.90 & 203.69& 4244.63 & 4266.07 &297.0& ~0.00\\
RC203 & 100 &4170.17& 4171.47& 4170.17& 463.80 &4170.17 & --& 368.40 &4170.17 &4177.62 & 196.96& 4171.90 & 4187.56 & 324.7& ~0.04\\
RC204 & 100 &4087.11& 4087.11& 4087.11& 347.40 &4087.11 & --& 328.20 &4087.11 &4094.04 & 171.97& 4087.11 & 4098.03 &315.7& ~0.00\\
RC205 & 100 &4291.93& 4291.93&4295.41& 327.60 &4293.73 & --& 251.40 &4291.93 &4294.56 & 197.21& 4295.33 & 4306.24 & 393.6& ~0.08\\
RC206 & 100 &4251.88& 4251.88& 4253.57& 307.20 &4251.88 & --& 256.20 &4251.88 &4257.78 &207.92& 4251.88 & 4259.31 & 367.3& ~0.00\\
RC207 & 100 &4182.44& 4185.98& 4186.43& 289.20 &4182.44 & --& 338.40 &4185.98 &4188.14 & 192.60& 4185.98 & 4192.18 & 307.2& ~0.08\\
RC208 & 100 &4075.04& 4075.04& 4076.27& 244.80 &4075.04 & --& 318.60 &4075.04&4077.57 & 166.98& 4076.74 & 4081.47 & 311.0& ~0.04\\
\bottomrule
\end{tabular}
  \end{threeparttable}
\end{sideways}
\vfill
\restoregeometry
\endgroup
 \clearpage

\subsection{FSMTWTF Instances}\label{sec:details_FSMTWTF}

\begingroup
\captionsetup{width=\textwidth}
\renewcommand\arraystretch{0.6}
{\footnotesize
\begin{longtable}{@{\extracolsep{\fill}} lll lllrrrrr rr}
\caption{Detailed results on the FSMTWTF instances (fleet A).\label{tab:FSMTWTF}}\\
\toprule
\multirow{2}{1.4cm}{Instance}  & \multirow{2}{0.8cm}{$|V_C|$} & \multirow{2}{0.8cm}{$|V_F|$} & \multicolumn{6}{c}{ALNS} & \multicolumn{2}{c}{FSMTW}\\
\cline{4-9}\cline{10-11}\rule{0pt}{2.5ex}  
  && & $z*$ & $\overline{z}$ & $\overline{\text{T}}$  &\#r&\#f&\#c  &BKS &Gap[\%] \\
\midrule
C101 &100& 10   &7085.43& 7086.00& 301.2 & 19 &1 &1 & 7093.45 &-0.11\\
C102 &100& 10   &7076.65 & 7076.96 & 270.5 & 19 &1 &1& 7080.17 & -0.05\\
C103 &100& 10   &7075.66 & 7076.23& 295.1 & 19 &1 &1 &7079.21& -0.05\\
C104 &100& 10   &7071.20 & 7073.33& 276.4 & 19 &1 &1 &7075.06&-0.05\\
C105 &100& 10   &7085.43 & 7085.78& 253.2 & 19 &1 &1&7093.45&-0.11\\
C106 &100& 10   &7081.30 & 7081.51& 233.6 & 19 &1 &1 &7083.87 & -0.04 \\
C107 &100& 10   &7082.95 & 7083.25& 224.2 & 19 &1 &1 &7084.61& -0.02   \\
C108 &100& 10   &7077.39 & 7077.80& 286.6 & 19 &1 &1  &7079.66& -0.03 \\
C109 &100& 10   &7075.03 & 7076.00& 244.9 & 19 &1 &1  &7077.30& -0.03 \\
\midrule 
R101 &100& 10   &4230.29& 4244.20& 310.6  & 19 &4 &20 & 4314.36&-1.95 \\
R102 &100& 10   &4142.25& 4160.42& 336.7 & 19 &3 &14 & 4166.28& -0.58\\
R103 &100& 10   &3994.49& 4008.85& 324.6 & 19 &4 &9& 4024.14&  -0.74\\
R104 &100& 10   &3936.00& 3945.02& 281.3 & 19 &1 &1 & 3936.40& -0.01 \\
R105 &100& 10   &4096.17& 4112.08& 290.5 & 20 &4 &11&  4122.50&  -0.64\\
R106 &100& 10   &4026.28& 4039.20& 322.6 & 19 &4 &14& 4047.34& -0.52 \\
R107 &100& 10   &3964.29& 3980.89& 304.9 & 19 &3 &7 & 3970.51& -0.16\\
R108 &100& 10   &3926.31& 3940.01& 319.2 & 19 &2 &3& 3928.12 & -0.05\\
R109 &100& 10   &3996.48& 4015.36& 304.0 & 19 &3 &11& 4011.97& -0.39\\
R110 &100& 10   &3959.76& 3968.13& 314.9 & 19 &2 &5 &  3960.55& -0.02 \\
R111 &100& 10   &3963.68& 3979.37& 333.1 & 19 &3 &8 & 3964.99&-0.03\\
R112 &100& 10   &3920.55& 3933.73& 325.3 &19 &1 &1 &  3917.88 & ~0.07\\
\midrule 
RC101 &100& 10   &5076.28& 5089.05& 340.7 & 17 &4 &19 &5150.86 &-1.45\\
RC102 &100& 10   &4929.53& 4944.06& 406.8 & 16 &3 &13 & 4974.82& -0.91 \\
RC103 &100& 10   &4799.78& 4815.85& 403.9 & 14 &2 &5&  4804.61& -0.10 \\
RC104 &100& 10   &4718.12& 4728.29& 399.4 & 12 &0 &0 &  4717.63 &~0.01\\
RC105 &100& 10   &4961.39& 4974.60& 389.4 & 16 & 2& 10 & 5035.35& -1.47\\
RC106 &100& 10   &4901.37& 4910.90& 414.5 & 16 & 3& 9 & 4919.43& -0.37\\
RC107 &100& 10   &4784.00& 4797.72& 432.7 &16 & 2& 6& 4787.59& -0.07\\
RC108 &100& 10   &4708.85& 4716.94& 414.4 & 15& 0& 0& 4708.85& ~0.00\\
\midrule 
C201 &100& 10   &5695.02& 5695.02& 427.2 & 5 &0 &0 &5695.02 & ~0.00\\
C202 &100& 10   &5685.24& 5685.24& 345.5 & 5 &0 &0 &5685.24& ~0.00\\
C203 &100& 10   &5681.55& 5698.11& 378.5 & 5 &0 &0 &5681.55& ~0.00\\
C204 &100& 10   &5677.66& 5677.66& 372.8 & 5 &0 &0 &5677.66&   ~0.00\\
C205 &100& 10   &5691.36& 5703.20& 415.9 & 5 &0 &0 &5691.36& ~0.00\\
C206 &100& 10   &5689.32& 5689.71& 414.3 & 5 &0 &0 &5689.32& ~0.00\\
C207 &100& 10   &5687.35& 5693.33& 422.1 & 5 &0 &0  &5687.35& ~0.00\\
C208 &100& 10   &5686.50& 5695.73& 391.8 & 5 &0 &0  &5686.50& ~0.00\\
\midrule 
R201 &100& 10   &3426.73& 3454.40& 427.8 & 5 &1& 4 & 3446.78 & -0.58\\
R202 &100& 10   &3295.10& 3301.80& 430.4 & 5 &1 &3 & 3297.42& -0.07  \\
R203 &100& 10   &3128.06& 3130.58& 426.1 & 5 &1 &5 & 3141.09& -0.41 \\
R204 &100& 10   &3014.24& 3054.73& 386.1 & 5&1 &4 &3018.14 &  -0.13\\
R205 &100& 10   &3220.81& 3227.40& 443.4 & 5 &0 &0& 3218.97& ~0.06  \\
R206 &100& 10   &3146.34& 3152.83& 437.6 & 5 &0 &0&  3146.34 & ~0.00 \\
R207 &100& 10   &3074.01& 3078.19& 410.9 & 5 &1 &4 & 3077.36& -0.11  \\
R208 &100& 10   &2998.70& 3001.31& 420.8 & 5 &0 &0&  2997.24& ~0.05 \\
R209 &100& 10   &3122.42& 3124.04& 428.9 & 5 &0 &0 & 3119.56& ~0.09  \\
R210 &100& 10   &3173.14& 3178.09& 442.4 & 5 &1 &3 &  3170.41& ~0.09 \\
R211 &100& 10   &3019.93& 3023.70& 452.9 & 5 &0 &0& 3019.93& ~0.00  \\
\midrule 
RC201 &100& 10   &4377.28& 4384.76& 414.6 & 12 &1 &10&  4374.09& ~0.07 \\
RC202 &100& 10   &4249.62& 4265.53& 432.6 & 13 &0 &0 & 4244.63& ~0.12  \\
RC203 &100& 10   &4174.00& 4178.18& 453.6 & 11 &0 &0 & 4170.17& ~0.09 \\
RC204 &100& 10   &4087.11& 4093.10& 438.5 & 9 &0 &0 &  4087.11& ~0.00\\
RC205 &100& 10   &4294.15& 4305.14& 419.3 & 12 &0&0 & 4291.93& ~0.05 \\
RC206 &100& 10   &4261.20& 4274.22& 446.4 & 11 &0 &0& 4251.88& ~0.22\\
RC207 &100& 10   &4185.98& 4208.76& 415.3 & 12& 0 &0 & 4182.44& ~0.08  \\
RC208 &100& 10   &4076.74& 4079.49& 421.9 & 9 &0 &0 & 4075.04 & ~0.04  \\
\bottomrule
\end{longtable}
}
\endgroup

\clearpage
\subsection{Additional Real-World-Instance Results}\label{sec:Real-World-Instance Results Figures}\label{sec:real-world-figs}

\begin{figure}[!htb]
\footnotesize
\centering
\includegraphics[width=\textwidth,keepaspectratio]{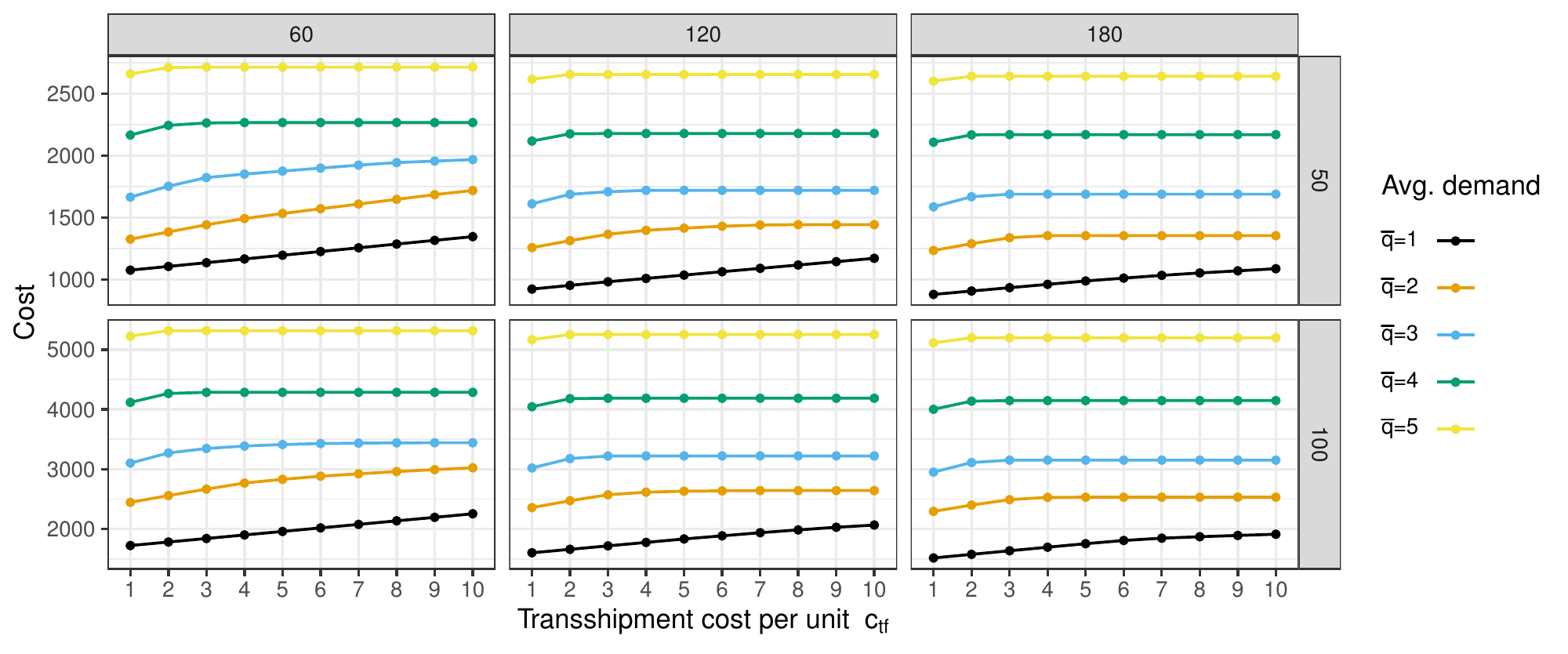}
\caption{Analysis of the impact of time window length $\kappa$ (columns), number of customers $|V_C|$, demand $\bar{q}$, and transshipment costs per unit $c_{tf}$ on the costs when using a homogeneous fleet.}\label{fig:ucc_cost_hom}
\end{figure}

\begin{figure}[!htb]
\footnotesize
\centering
\includegraphics[width=\textwidth,keepaspectratio]{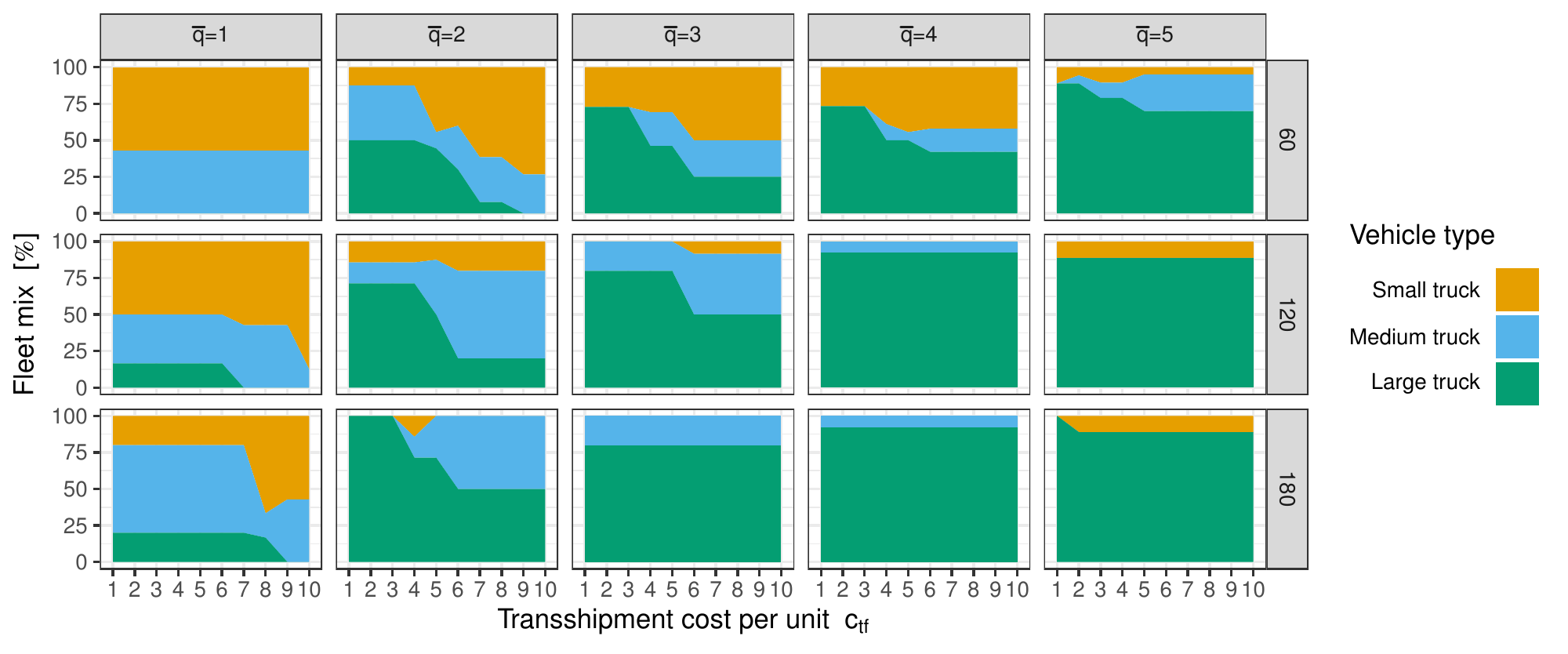}
\caption{Analysis of the impact of time window length $\kappa$ (rows), demand $\bar{q}$, and transshipment costs per unit $c_{tf}$ on the fleet mix for $|V_C|=100$.}\label{fig:fleet_mix_100}
\end{figure}

\end{document}